\documentclass{article}
\usepackage[a4paper,margin=0.5in,footskip=0.25in]{geometry}
\usepackage[backend=biber, style=numeric-comp, sorting=none]{biblatex}
\usepackage{amsmath,amssymb,amsfonts,dsfont,xspace,graphicx,relsize,bm,mathtools,xcolor,amsthm,soul}
\usepackage{graphicx}
\usepackage{dcolumn}
\usepackage{xcolor}
\usepackage{bbm}
\usepackage{physics}
\usepackage{ulem}
\usepackage{hyperref}
\hypersetup{
	colorlinks = true,
	urlcolor   = blue,
	linkcolor  = blue,
	citecolor  = magenta
}
\usepackage{algorithm}
\usepackage{algpseudocode}
\usepackage{float}
\usepackage[shortlabels]{enumitem}
\usepackage{subfigure}
\usepackage{orcidlink}
\usepackage{braket}
\usepackage{subcaption}
\usepackage{authblk}

\begin{document}
\title{Implementation of Leaking Quantum Walks on a Photonic Processor}

\author[1]{E. Stefanutti}
\author[2]{J. Philipps}
\author[2]{J. Buetow}
\author[2]{A. Guidara}
\author[1]{M. Nuvoli}
\author[1,3]{A. Chiuri}
\author[1]{L. Sansoni}
\affil[1]{ENEA - Nuclear Department, Via E. Fermi 45, 00100 Frascati, Italy}
\affil[2]{QuiX Quantum B.V., 7521 AN Enschede, The Netherlands}
\affil[3]{INFN Sezione Roma Tre, Via della Vasca Navale, 84, 00146 Rome, Italy}
\date{}                     
\setcounter{Maxaffil}{0}
\renewcommand\Affilfont{\itshape\small}

\maketitle

\abstract{Quantum walks (QWs) represent pillars of quantum dynamics and information processing. They provide a powerful framework for simulating quantum transport, designing search algorithms, and enabling universal quantum computation. Several physical platforms have been employed for their implementation, such as trapped atoms and ions, nuclear magnetic resonance systems, and photonic quantum architectures either in bulk optics or waveguide structures and fiber-loop networks. Here we focus on the most promising and versatile approach, that is photonic integrated circuits.
In this work, we review how the employment of this versatile experimental platform has allowed to explore several phenomena related to QW-based protocols as, for instance, the evolution in presence of different kinds of noise. In this landscape, to the best of our knowledge, few examples report on the introduction of absorbing centers and their effects on the coherence of the dynamics. Here we present and discuss the results related to absorbing boundaries in QWs obtained through theoretical simulations and experiments conducted with the universal photonic quantum processors realized by Quix Quantum. We analyze how localized absorption along one lattice edge affects the walker dynamics depending on both the leakage probability and the initial injection site. Our results show that the presence of controlled losses modifies interference patterns and coherence, without fully destroying quantum features and providing an effective resource for engineering on-chip QWs and simulating open quantum systems.\\
\noindent \textbf{Keywords}: photonic quantum walk; leaking probability; absorbing boundary; confinement; photonic integrated circuits}

\section{Introduction}
Quantum walks (QWs) have emerged as a versatile framework in quantum information science, providing both a fundamental model of coherent quantum dynamics and a practical resource for information processing tasks \cite{Aharonov1993,Kempe2003,VenegasAndraca2012,Kadian2021}. Owing to their intrinsically quantum features—such as superposition, interference, and entanglement—QWs exhibit dynamics that differ markedly from classical random walks, enabling algorithmic speedups in search and graph-based problems \cite{Childs2009,AMBAINIS2003,Shenvi2003}. Beyond quantum algorithms, QWs constitute a powerful platform for quantum simulation, allowing controlled investigations of transport, disorder-induced localization, and topological phases in complex networks \cite{AspuruGuzik2012,Kitagawa2010,Broome2010}.
QWs have been implemented in various platforms \cite{Wang2013_book}, ranging from cold atoms \cite{Dadras2018,Dadras2019,Clark2021} to superconducting devices \cite{Gong2021} and photonic setups \cite{Peruzzo2010,Schreiber2010,Sansoni2012}.
Optical implementations of QWs have enabled the investigation of genuinely multi-particle effects, including the role of particle statistics in quantum diffusion \cite{Sansoni2012} as well as non-trivial three-photon interference phenomena \cite{Zhou2024}.
Moreover, QWs provide a versatile testbed for studying the impact of noise on quantum coherence: controlled disorder has been introduced to observe Anderson localization in the presence of static noise \cite{Crespi2013,DeNicola2014,Schreiber2011}, while QWs on multidimensional lattices have been experimentally realized using both fiber-loop architectures \cite{Schreiber2012} and femtosecond-laser-written waveguide circuits \cite{Tang2018}.
Among the various experimental platforms, integrated photonic circuits stand out as particularly promising, offering high phase stability, intrinsic resilience to noise and decoherence, and clear prospects for scalability \cite{Wang2019}. Integrated photonic devices for QWs have also proven ideal for the implementation of boson sampling protocols \cite{Tillmann2013,Crespi2013bis,Spagnolo2014,Zhong2019,Hoch2022,Anguita2025arx}, providing an exemplary study of how quantum supremacy can be achieved \cite{Aaronson2011,Lund2017,Hamilton2017}.

In real photonic quantum systems, particle losses are unavoidable due to the non-ideal efficiencies of the network components and measurements apparatuses, often representing a major limitation for scaling up quantum applications. Nonetheless, the inclusion of absorbing sites within a QW remains a scenario that has received relatively little attention and has not yet been fully explored. In this regard, a limited number of numerical studies \cite{Bach2004,Stefanak2008,Wang2016,Kulinski2020,Ammara2025arx} and a few experimental implementations \cite{Nitsche2018sciad,Pegoraro2023} have been reported, showing how some recursive behavior is observed in the presence of specific absorbing sites. However, many aspects of this problem remain largely unexplored, even though the introduction of controlled losses appears to be a promising approach for probing the dynamics of open quantum systems.

In this work, we address this gap by presenting both numerical simulations and experimental results for a QW in the presence of an absorbing boundary. Specifically, we investigate a walk on a finite lattice featuring a partially absorbing (leaking) boundary of tunable strength and a fully reflective boundary. We observe that the absorbing boundary significantly alters the walker’s evolution, with the magnitude of this effect increasing with the absorption strength.

The employed finite system represents an effective minimal model for mimicking energy transport in complex networks, such as those found in biological systems \cite{Biggerstaff16,mohs08jcp}. Here an interesting example is represented by the well-known Fenna-Matthews-Olson (FMO) complex of green sulfur bacteria which is characterized by 7 sites and essentially acts as a molecular wire, transferring excitation energy while showing long-lived quantum coherence \cite{Engel2007}. Indeed, the FMO complex has been proposed as a dedicated computational device \cite{Engel2007}, as excitons are able to explore many states simultaneously and efficiently select the correct answer.
Our study expands what has been already demonstrated employing the QW approach to introduce further complexity in this dynamics. Precisely, we introduce possible effects due to sites coupling with the environment resulting into energy absorption and decoherence.  
On the other hand, with this manuscript we explore a field with multiple possible applications, as controllable decoherence permits photonic implementations of quantum-computational methods that take advantage of decoherence  \cite{Kendon2003,KENDON_2007,Verstraete2009}. In this context, the optimization of suitable transport processes ---whether coherent or noise-assisted--- may require the engineering of dedicated quantum systems.

\section{Materials and Methods}

\subsection{Theoretical background and simulations}
In this work, we investigated the effects of mode-dependent particle losses within a noise-free discrete-time quantum walk (DTQW) \cite{Venegas2012} involving single photons. To this end, we considered a QW evolving over $N$ temporal steps, where homogeneous losses were introduced in a selected propagation mode located at the edge of the lattice.
Our numerical approach is based on a coined DTQW model, where the time evolution of the photonic walker is described as the the result of two operations: a coin toss $\hat{C}$ followed by a conditional displacement $\hat{S}$. Accordingly, the dynamics is governed by the evolution operator $\hat{U} = \hat{S}(\hat{C}\otimes \hat{I})$, acting on the composite Hilbert space $\mathcal{H} = \mathcal{H_C} \otimes \mathcal{H_S}$. Here $\mathcal{H_S}$ denotes the position subspace, spanned by orthonormal site states $\ket{x}$ along a finite one-dimensional lattice, while $\mathcal{H_C}$ is the internal two-dimensional coin subspace, spanned by $\{\ket{L}, \ket{R}\}$, encoding the left/right direction of propagation for the next hop. The generic state of the system can thus be written as 
\begin{equation}
\ket{\Psi} = \sum_{x \in \mathbb{Z}} (\alpha_{L,x}\ket{L} + \alpha_{R,x}\ket{R}) \otimes \ket{x}
\end{equation}
where $|\alpha_{L,x}|^2$ and $|\alpha_{R,x}|^2$ represent the probabilities for the walker at site $x$ to move at the following step to the left or to the right, respectively, with the normalization condition for the coin state $|\alpha_{L,x}|^2+|\alpha_{R,x}|^2=1, \forall x$. The conditional displacement is realized by the shift operator: 
\begin{equation}
\hat{S} = \sum_{x} \ket{L}\bra{L} \otimes \ket{x-1}\bra{x} + \ket{R}\bra{R} \otimes \ket{x+1}\bra{x}
\end{equation}
which displaces the walker in a superposition of the position basis states according to its internal coin state. As a result, the state of the system at each time step depends recursively on the previous step, i.e. $\ket{\Psi(n)} = \hat{U} \ket{\Psi(n-1)}$, yielding after $n$ steps the coherent evolution $\ket{\Psi(n)} = \hat{U}^n \ket{\Psi(0)}$, where $\ket{\Psi(0)}$ denotes the initial state of the walker. 
Throughout this work, we adopted the Hadamard coin operator, $\hat{C} = \frac{1}{\sqrt{2}} \left( \begin{matrix} 1 && 1 \\ 1&&-1 \end{matrix} \right)$, which mimics a fair coin toss. 

Experimentally, such a coined DTQW can be realized on photonic platforms by injecting photons through a cascade of directional couplers acting as balanced beam splitters (BSs), arranged in a lattice of elementary cells forming Mach–Zehnder interferometers (MZIs). Each BS simultaneously implements both the coin and shift operators, since it splits the photon into left- and right- propagation path and shifts it accordingly.
In this scheme, each BS output represents a point in the space-time evolution of the QW: at a given step, the walker is described by the state $\ket{x,n}$, where $x$ labels the position of the walker along the line corresponding to the spatial mode and $n$ denotes the discrete time ($0 \leq n \leq N, n \in \mathbb{N}$). In this configuration, $2M$ ($M \in \mathbb N$) defines the number of lattice sites available at each time step, while $M$ is the maximum number of MZI units per step. Thus the variable $x$ takes discrete values given by $-x_{max}+(m-1)$, where $x_{max}=M-0.5$ and $m$ is an integer index labeling the injection site, with $m\in(0,2M]$. 

\begin{figure}[ht]
    \centering
    \includegraphics[width=.7\linewidth]{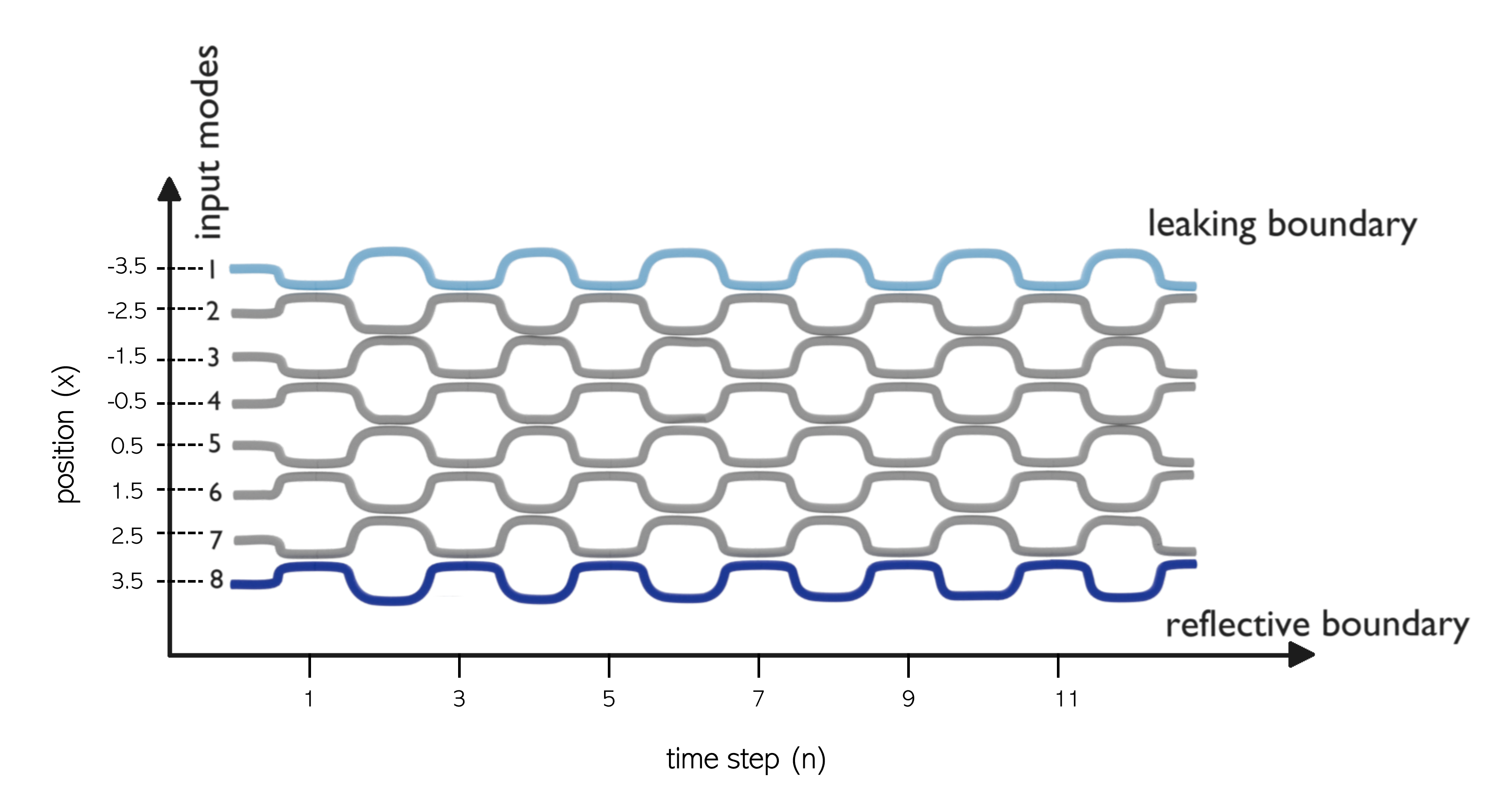}
    \caption{Scheme of the confined leaking QW with $2M=8$ modes, implemented as a cascade of MZ interferometers. The pale-blue waveguide at the upper edge of the lattice (input mode 1, corresponding to the position $x=-3.5$) represents the leaking boundary, where photon are partially lost. The dark-blue waveguide at the bottom edge (input mode 8, corresponding to $x=3.5$ along the position axis) depicts hard-confinement boundary, where photons are totally reflected back into the lattice. The internal gray waveguides (input modes 2-7) implement Hadamard coin operation with balanced BS. Light is injected from one of the input modes as described in the text.}
    \label{fig:Leaking_QW_scheme}
\end{figure}

Building upon our previous results on confined QWs \cite{Sansoni2025arxiv}, we extended the analysis by introducing an additional key aspect beyond spatial confinement induced by lattice edges, namely the presence of a leaking boundary. 
We impose asymmetric boundary conditions: one edge of the lattice enforces hard confinement, while the opposite edge acts as a homogeneous leaking boundary. In the former case, the walker is completely reflected at the edge and redirected back into the interior whenever a step would take it outside the lattice. In the latter case, reflection at the boundary is only partial, resulting in a finite probability for the wavefunction to leak out of the lattice, thereby introducing controlled, mode-dependent particle losses (Figure~\ref{fig:Leaking_QW_scheme}).
In order to model this behavior numerically, we constructed a $2M \times 2M$ block-diagonal matrix $M_{BS}$, whose diagonal blocks are represented by the BS transformation 
\begin{equation}\label{eq:BS_matrix}
\begin{pmatrix}
    \sqrt{1-r^2} & r\\
    -r &  \sqrt{1-r^2}
\end{pmatrix}    
\end{equation}
at each time step, with $0<r^2<1$. For the internal sites $m$ not corresponding to modes at the edges ($ 2 \leq m \leq M-1$), each block given by Equation~\ref{eq:BS_matrix} consisted of a $2\times2$ balanced BS transformation with transmissivity $r^2 = 1/2$, in order to implement the Hadamard coin. At the lattice edges, the balanced BS was replaced by an unbalanced transformation with transmissivity $r^2 \neq 0.5$. In particular, $r^2=0$ models a perfectly reflective boundary (lossless edge), whereas $r^2<0.5$ and $r^2>0.5$ model low- and high-leakage scenarios, respectively. 
At each time step, the block matrix structure accounted for the alternate pattern of an even and an odd number of BSs over subsequent time steps. Here $M_{BS}$ represented the transformation associated with the evolution operator $\hat{U}(n)$ at the $n$-th time step. Hence, the complete evolution of the QW after $n$ steps was obtained as the ordered product: $\ket{\Psi(n)} = \hat{U}(n) \, \hat{U}(n-1) \ldots \hat{U}(2) \, \hat{U}(1)$.

In a QW, the spatial probability profile of the walker after multiple time steps reflects nontrivial interference effects arising from the coherent superposition of the many possible propagation paths. To capture this behavior, we computed the output single-particle probability distributions across all spatial modes, $P(x;n) = |\langle x|\psi(x,n)\rangle|^2$, which represents the probability of finding the particle at position $x$ after $n$ steps, irrespectively of its internal coin state. In the presence of a leaking boundary, the dynamics is intrinsically non-unitary: since a fraction of the wavefunction is irreversibly lost due to the interaction with leaking edge, the total probability is no longer conserved throughout the time evolution. Therefore, the probability distributions here reported are referred to the population that remained confined within the lattice at each time step. 
We analyzed the propagation of single photons injected at different initial positions, both close to or far from the leaking boundary, in order to assess the role of the distance from the leaking edge for different leaking probabilities. 
As quantitative indicators for characterizing the walker dynamics, we computed the time evolution of mean position $\langle x \rangle$ and its variance $\sigma_n^2(x)$, defined as:
\begin{eqnarray}
    &&\langle x\rangle=\sum_{i=1}^8 p_ix_i\nonumber\\
    &&\sigma^2_n(x)=\langle x^2\rangle-\langle x\rangle^2=\sum_{i=1}^8 p_ix_i^2-\left(\sum_{i=1}^8 p_ix_i\right)^2
    \label{eq:variance}
\end{eqnarray}
where $x_i$ denotes the output position and $p_i$ the corresponding probability.
\subsection{Experimental setup}
The model described above is implemented in a photonic platform. A coherent attenuated beam is used as the walker, while the lattice with asymmetric boundaries is realized using a reconfigurable photonic processor (QuiX Quantum Alquor20) featuring 20 input and output ports \cite{Taballione2021,Taballione2023}. The processor is a fully programmable multiport interferometer capable of implementing arbitrary linear optical transformations over a space whose dimensionality is set by the number of available modes.
The interferometer is constructed as a mesh of 190 MZIs interferometers, each functioning as a tunable beam splitter \cite{Clements16}, thereby enabling independent control of the amplitudes and phases of the output signals. The photonic chip is fabricated using stoichiometric silicon nitride ($\mathrm{Si_3N_4}$) waveguides based on TripleX technology and is mounted on a water-cooled Peltier element to ensure thermal stability. Optical coupling to and from the chip is provided via FC/PC fiber connectors.
The processor exhibits an insertion loss of $(3.65 \pm 1.30)\,\mathrm{dB}$. Its performance is characterized by an average amplitude fidelity of $F = (98.8 \pm 0.3)$, evaluated over 100 Haar-random unitary matrices at a wavelength of $942\,\rm{nm}$. The fidelity is defined as $F = \frac{1}{d} Tr(|U^{\dagger}_{th} \cdot U_{exp}|)$, where $|U|$ denotes the element-wise absolute value and $d = 20$ is the number of ports \cite{wang09}. The fully packaged chip is mounted on a sub-mount to ensure mechanical stability and is electrically interfaced with a printed circuit board (PCB). Control of the device is provided through a Python-based software interface, which allows the user to specify the target transformation and accordingly program the phase shifts applied to the MZIs.

\begin{figure}[h]
    \centering
    \includegraphics[width=0.8\linewidth]{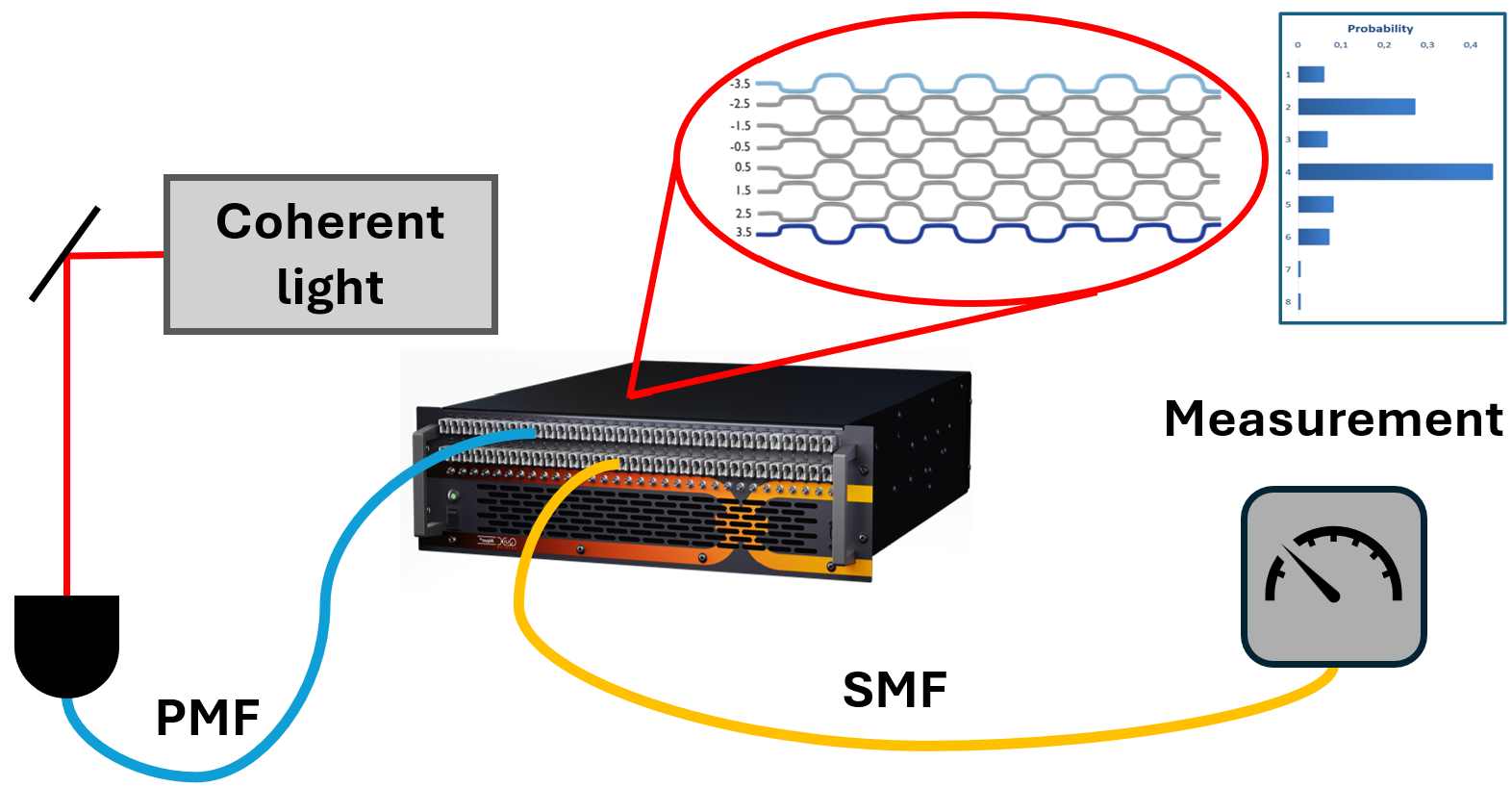}
    \caption{Scheme of the experimental setup. Attenuated coherent light plays the role of the walker which is injected into the tunable photonic processor via a polarization maintaining fiber (PMF). The processor implements the leaking QW unitary and the output is sent to the measurement stage through single mode fibers (SMF). The measurement apparatus consists of a photodiode used to retrieve the probability distributions (blue inset) across the output modes. Thanks to a computer-driven reconfigurability of the processor, we implemented various leaking configurations.
    }
    \label{fig:setup}
\end{figure}

A schematic of the experimental setup is shown in Figure~\ref{fig:setup}. Attenuated coherent light at a wavelength of $\lambda = 942\,\mathrm{nm}$ is injected into the selected input mode of the photonic processor via a polarization-maintaining fiber. The processor is programmed to implement the unitary transformation corresponding to the quantum walk with asymmetric boundaries under investigation.
The dimensionality of the multimode interferometer enables the realization of quantum walks of up to $N = 20$ steps; in the transverse direction, $2M = 8$ input and output ports are employed. The selected input mode is initially populated by injecting approximately $10\,\mu\mathrm{W}$ of optical power through a polarization-maintaining fiber.
For each $n$-step QW, with $4 \leq n \leq 20$, the appropriate unitary operator is applied and the output intensities are measured across all eight output channels. Detection is performed using photodiodes coupled via single-mode fibers, allowing reconstruction of the output probability distributions. To account for variations in transmission among the output fibers, a calibration procedure is carried out by applying the identity operation on the processor and using a fixed reference output. After this correction, the measured intensity distributions are normalized to unity.
\section{Results}

\subsection{Simulations}\label{sec:Results_sim}
In order to ensure that boundary effects significantly affected the dynamics, we considered a DTQW on a number of sites small compared to the characteristic diffusion scale of the walker. Specifically, we simulated a QW on a lattice of $2M = 8$ sites over up to $N = 100$ discrete time steps, thereby assessing the long-time behavior. The choice of $2M=8$ sites was also guided by the need to allow direct comparison between simulation results and experimental data.
We considered four distinct cases depending on the position of the site at which the walker is injected into the lattice: positions close to the leaking boundary (input waveguides 2 and 3, corresponding to the initial states $\ket{-2.5,0}$ and $\ket{-1.5,0}$, respectively), and positions close to the reflective boundary (input waveguides 6 and 7, corresponding to the initial states $\ket{1.5,0}$ and $\ket{2.5,0}$, respectively). 
In our simulations, we adopted $r^2=0.2$ for the low-leakage regime and $r^2=0.8$ for the high-leakage case. In the present model, particle losses at the leaking boundary are assumed to be homogeneous and time-independent, i.e., the leakage probability is the same at each time step.
In Figure~\ref{fig:Sim_100steps_cfr_Input}, the walker mean position and its variance, evaluated over 100 steps, are compared for different input positions and for both the low-loss and high-loss regimes. A more detailed comparison between weak- and strong-leakage dynamics for each input site, including lossless and fully absorbing regimes as reference cases, is reported in Appendix~\ref{AppendixA_Results_Results}.

\begin{figure}[ht]
\centering
\begin{minipage}{0.48\textwidth}
    \centering
    \includegraphics[width=\linewidth]{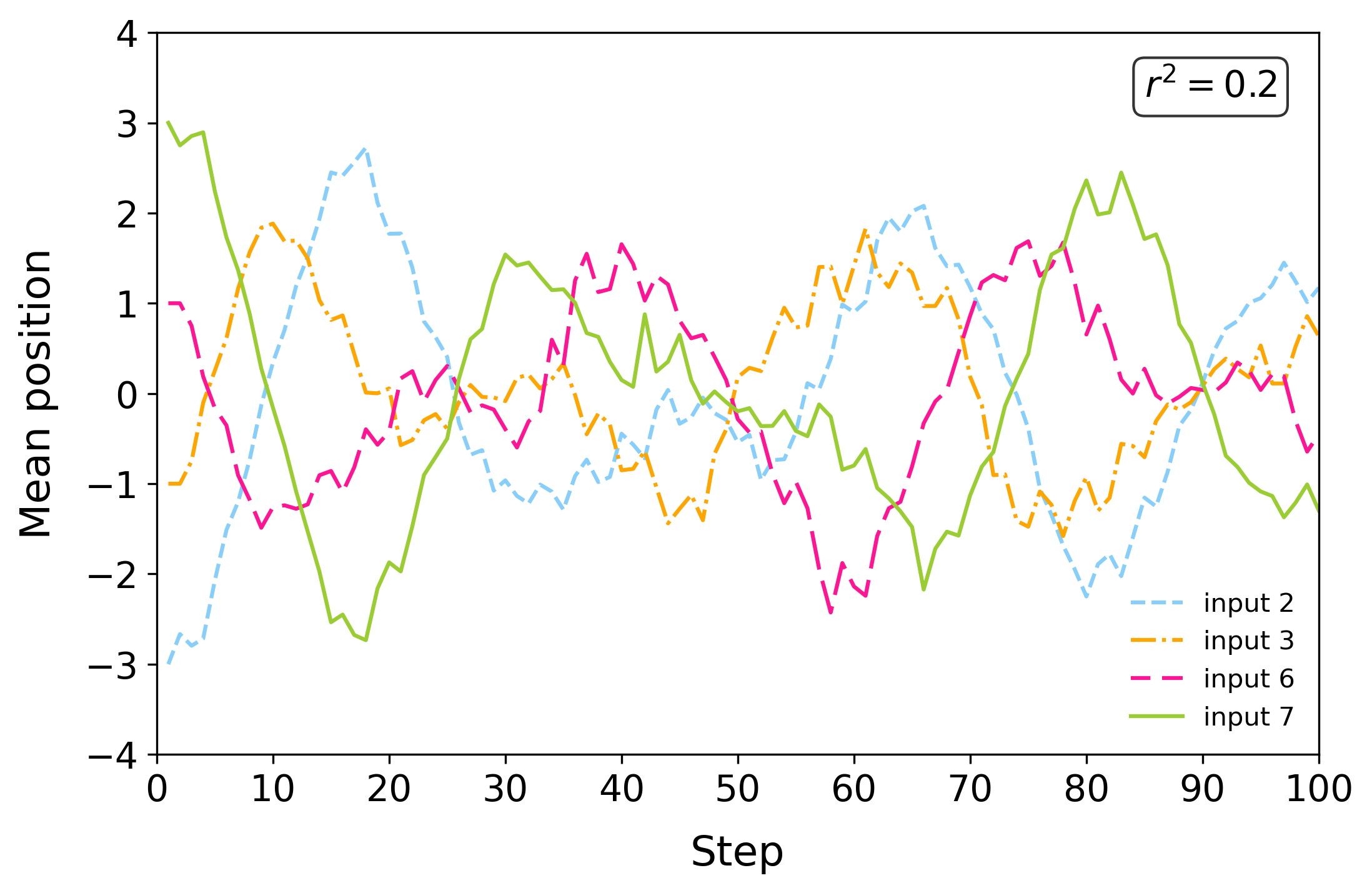}
    (a)
\end{minipage}\hfill
\begin{minipage}{0.48\textwidth}
    \centering
    \includegraphics[width=\linewidth]{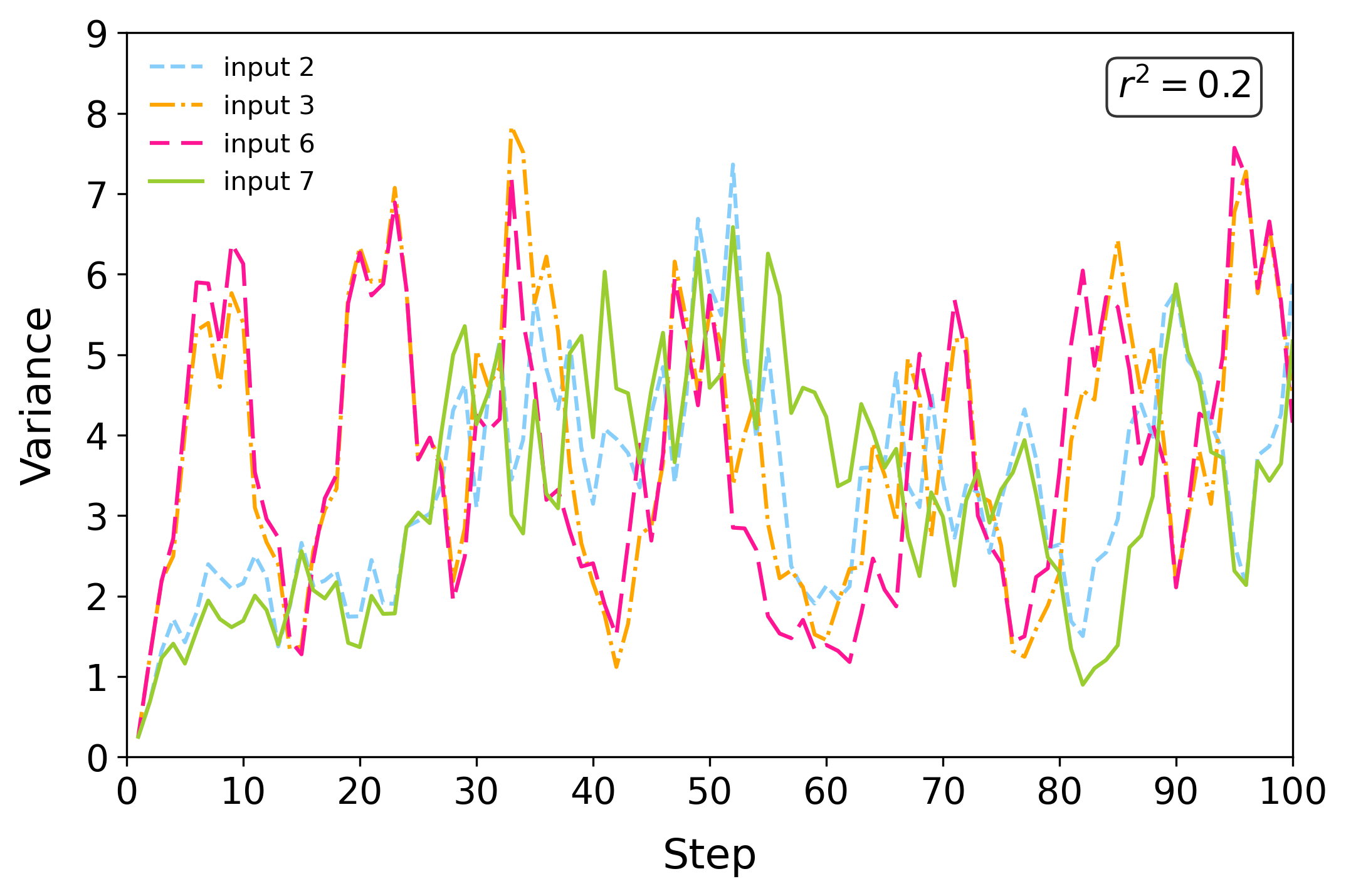}
    (b)
\end{minipage}
\vspace{0.3cm}
\begin{minipage}{0.48\textwidth}
    \centering
    \includegraphics[width=\linewidth]{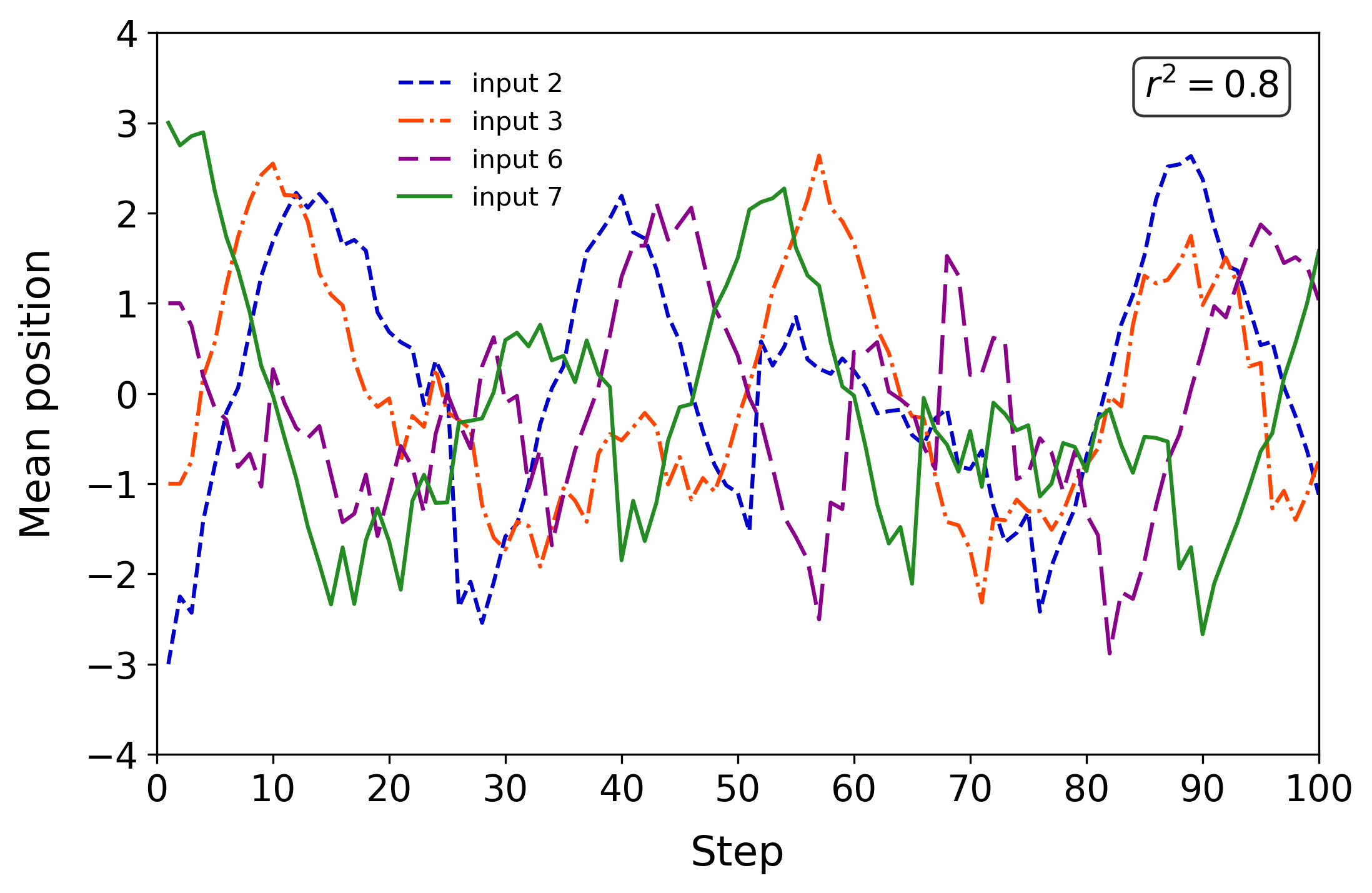}
    (c)
\end{minipage}\hfill
\begin{minipage}{0.48\textwidth}
    \centering
    \includegraphics[width=\linewidth]{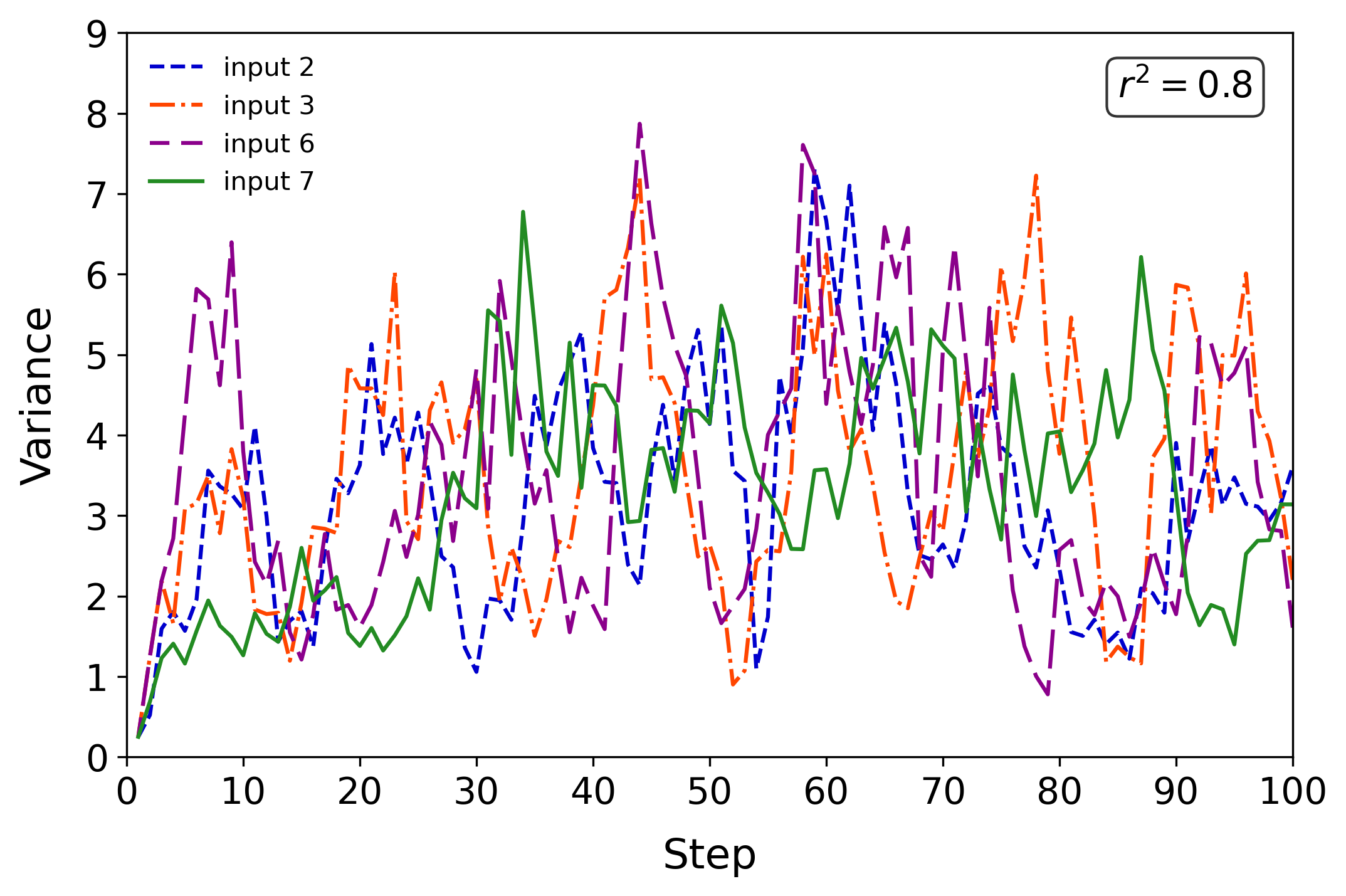}
    (d)
\end{minipage}
\caption{Simulated mean position (left column, panels (a) and (c)) and variance of the mean position (right column, panels (b) and (d)) as a function of the number of steps: comparison between the four selected input modes for leaking probability $r^2 = 0.2$ (panels (a) and (b)) and $r^2=0.8$ (panels (c) and (d)), corresponding to low- and high-leakage regime, respectively.}
\label{fig:Sim_100steps_cfr_Input}
\end{figure}

Due to the confinement, the walker propagation across the lattice follows oscillatory patterns, resulting from a sequence of total or partial reflections at the edges of the accessible region. This behavior is captured by the time evolution of both the mean position and its variance (Figure~\ref{fig:Sim_100steps_cfr_Input}\,(a)-(d)).
In the low-loss regime ($r^2=0.2$), the dynamics appears to be only weakly affected by the presence of the leaking boundary and, as shown in Figure~\ref{fig:Sim_100steps_cfr_Input}\,(a), the mean position depends on the injection site: when the walker enters the lattice from sites closer to one of the boundaries (input sites 2 and 7, cyan dashed and green solid lines), it propagates towards the opposite edge, whereas for more internal injection sites (3 and 6, orange dash-dotted and long-dashed pink lines) the mean position exhibits oscillations of smaller amplitude around the origin of the position axis. Nevertheless, the symmetry associated with the spatial coordinates of the input sites is largely preserved: opposite injection sites (2 and 7, or 3 and 6) share essentially the same temporal evolution, but with opposite phase.   
Likewise, the time evolution of the variance of the walker’s mean position exhibits an oscillatory behavior (Figure~\ref{fig:Sim_100steps_cfr_Input}\,(b)). For each pair of symmetric input sites, the variance follows the same qualitative temporal evolution, indicating that the spreading and partial refocusing of the wavefunction are similarly affected by the lattice boundaries when the leaking probability is low. In particular, when the walker is injected close to either the leaking or the reflecting edge (cyan dashed and green solid lines), the initial spreading remains limited and the variance stays below 3 for approximately the first 25 steps, before gradually increasing at longer times while preserving fluctuations of large amplitude and high frequency. On the other hand, for more internal injection sites the variance displays pronounced oscillations for the earliest steps, corresponding to alternative phases of spreading and relocalization of the wavefunction, occurring with a characteristic period of about 15 steps (orange dash-dotted and long-dashed pink lines).

By contrast, in the high-loss regime ($r^2=0.8$), the dynamics of both the mean position and its variance, although still retaining oscillatory features, departs from the regularities and phase relations identified in the low-loss case (Figure~\ref{fig:Sim_100steps_cfr_Input}\,(c),(d)). In particular, symmetric injection sites no longer display identical dynamics, and both the amplitude and frequency of the oscillations are modified, indicating a stronger impact of losses on the walker propagation.  
To further clarify the role of increasing losses on the system dynamics, it is helpful to compare the time evolution of the walker's mean position and variance for different leaking probabilities and injection sites. The full set of long-time numerical results is presented in Appendix~\ref{AppendixA_Results}.

As expected, the strongest sensitivity to losses is observed for input site 2, which is the closest to the leaking boundary (Figure~\ref{fig:Sim_100steps_cfr_T}\,(a)-(b)). As long as the losses remain weak ($r^2=0.2$, cyan solid line), the dynamics closely follow that of the lossless case ($r^2=0$, gray dotted line), with few differences limited to their relative amplitudes up to approximately 40-50 steps. However, when the losses become strong ($r^2=0.8$, blue dashed line), deviations appear already at early times, with a faster propagation of the mean position (for $r^2=0.2$ the mean position reaches the first oscillation peak within the first 18 steps, whereas for $r^2=0.8$ it requires only 12 steps to reach the first maximum mean displacement) and an enhanced spreading of the wavefunction, as evidenced by the larger variance.  
For injection sites progressively farther from the leaking boundary (input sites 3, 6, and 7), the impact of losses on the walker's dynamics appears to be reduced (Figure~\ref{fig:Sim_100steps_cfr_T}\,(c)-(h)). In these configurations, the mean position and variance exhibit similar trends across different leaking regimes for a large number of time steps, with only minor differences in amplitude and phase. This indicates that, when the walker is injected sufficiently far from the lossy edge, the dynamics remain largely governed by coherent propagation within the lattice.  
In the following, we focus in particular on the initial $20$ time steps, for which a direct comparison with experimental observations was feasible.In order to ensure that boundary effects significantly affected the dynamics, we considered a DTQW on a number of sites small compared to the characteristic diffusion scale of the walker. Specifically, we simulated a QW on a lattice of $2M = 8$ sites over up to $N = 100$ discrete time steps, thereby assessing the long-time behavior. The choice of $2M=8$ sites was also guided by the need to allow direct comparison between simulation results and experimental data.
We considered four distinct cases depending on the position of the site at which the walker is injected into the lattice: positions close to the leaking boundary (input waveguides 2 and 3, corresponding to the initial states $\ket{-2.5,0}$ and $\ket{-1.5,0}$, respectively), and positions close to the reflective boundary (input waveguides 6 and 7, corresponding to the initial states $\ket{1.5,0}$ and $\ket{2.5,0}$, respectively). 
In our simulations, we adopted $r^2=0.2$ for the low-leakage regime and $r^2=0.8$ for the high-leakage case. In the present model, particle losses at the leaking boundary are assumed to be homogeneous and time-independent, i.e., the leakage probability is the same at each time step.
In Figure~\ref{fig:Sim_100steps_cfr_Input}, the walker mean position and its variance, evaluated over 100 steps, are compared for different input positions and for both the low-loss and high-loss regimes. A more detailed comparison between weak- and strong-leakage dynamics for each input site, including lossless and fully absorbing regimes as reference cases, is reported in Appendix~\ref{AppendixA_Results_Results}.

\begin{figure}[ht]
\centering
\begin{minipage}{0.48\textwidth}
    \centering
    \includegraphics[width=\linewidth]{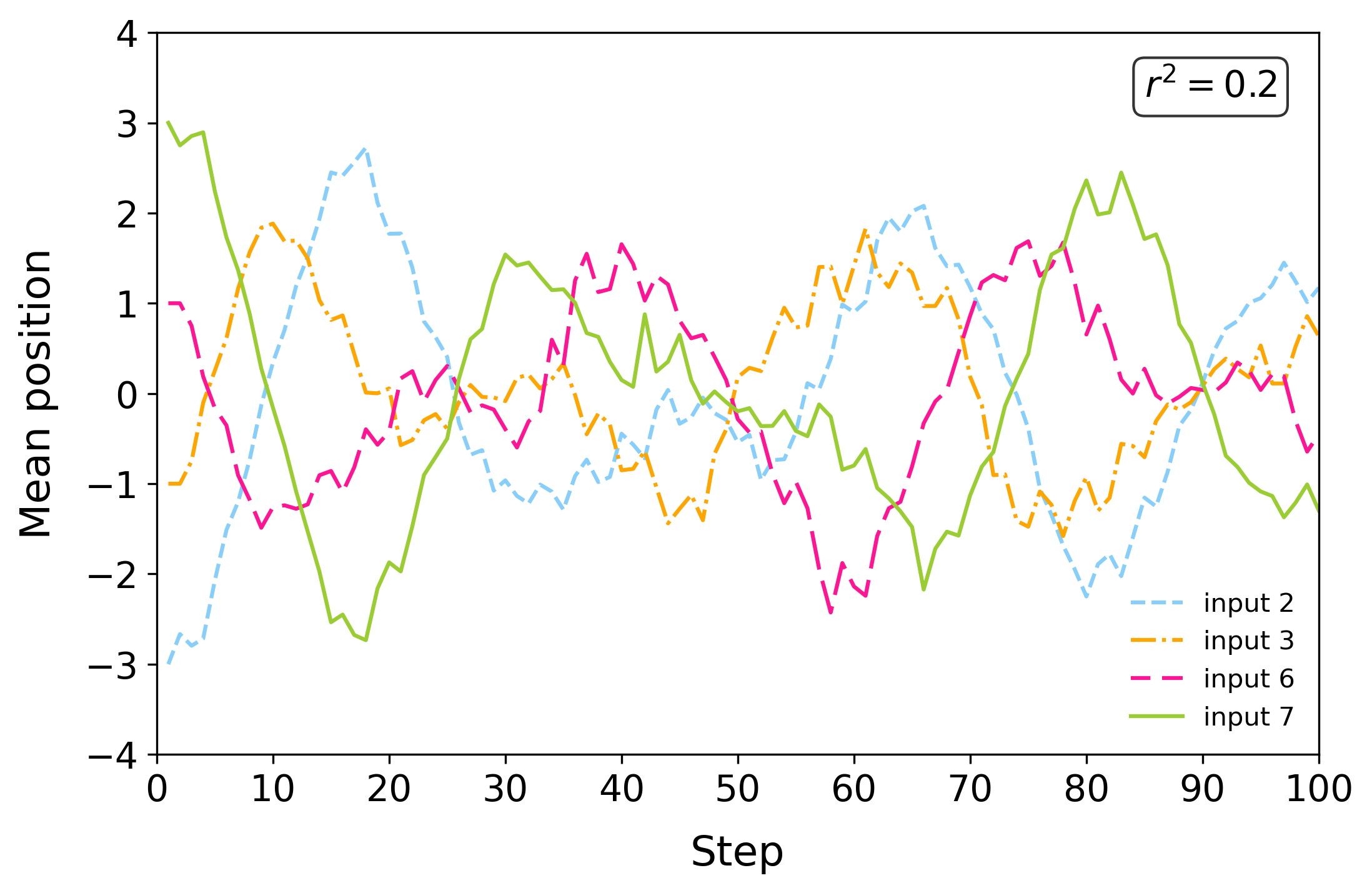}
    (a)
\end{minipage}\hfill
\begin{minipage}{0.48\textwidth}
    \centering
    \includegraphics[width=\linewidth]{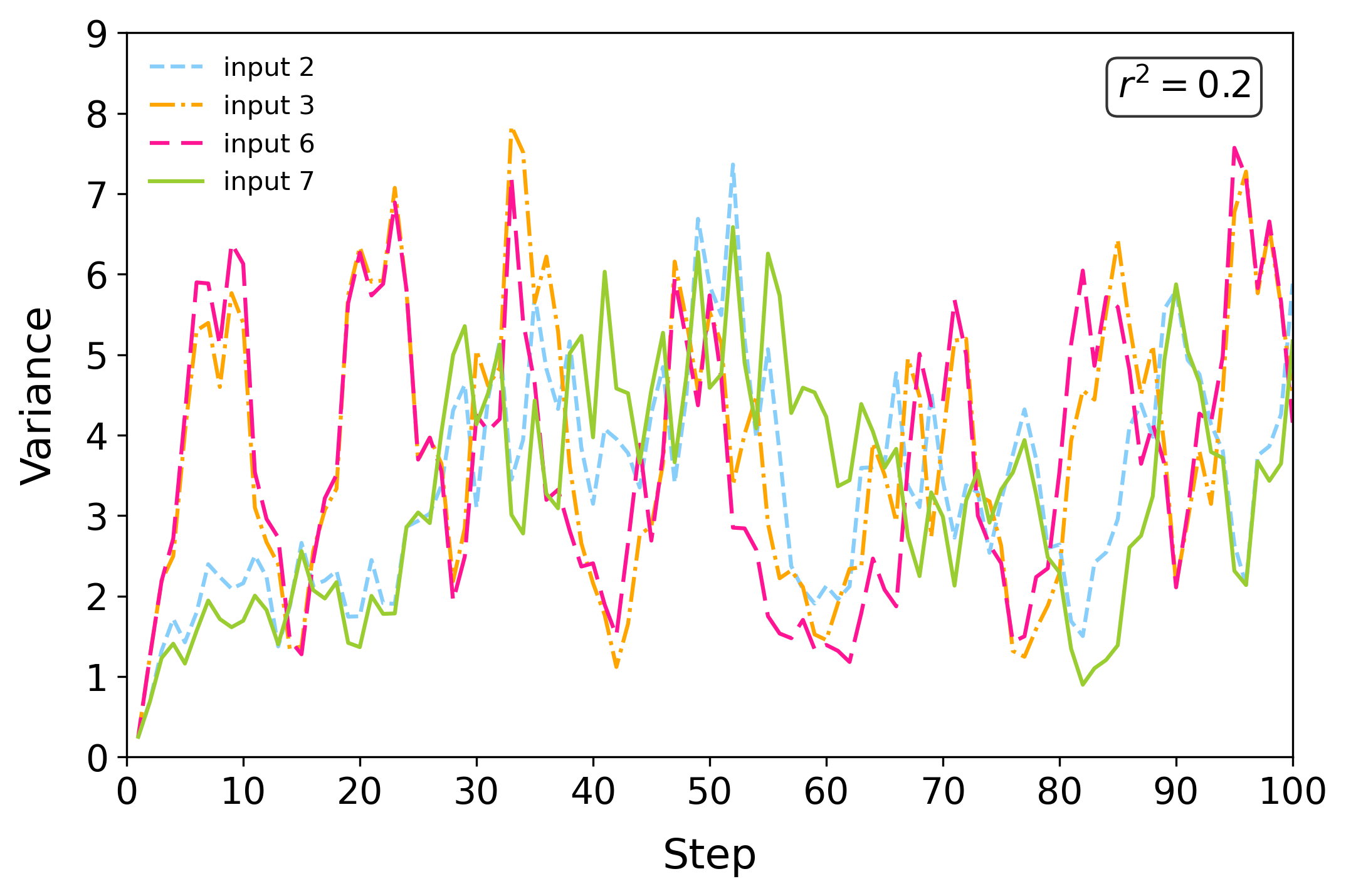}
    (b)
\end{minipage}
\vspace{0.3cm}
\begin{minipage}{0.48\textwidth}
    \centering
    \includegraphics[width=\linewidth]{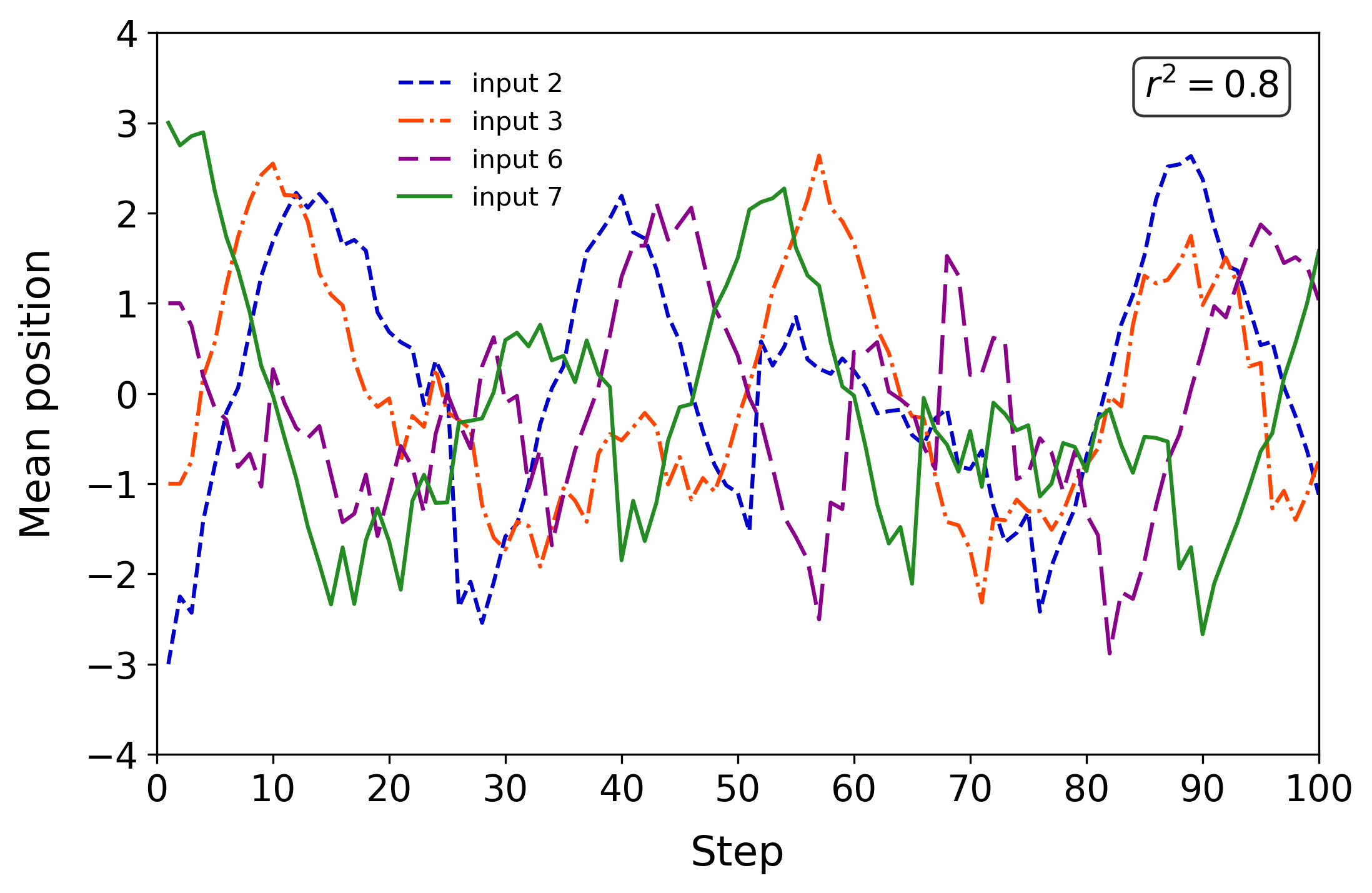}
    (c)
\end{minipage}\hfill
\begin{minipage}{0.48\textwidth}
    \centering
    \includegraphics[width=\linewidth]{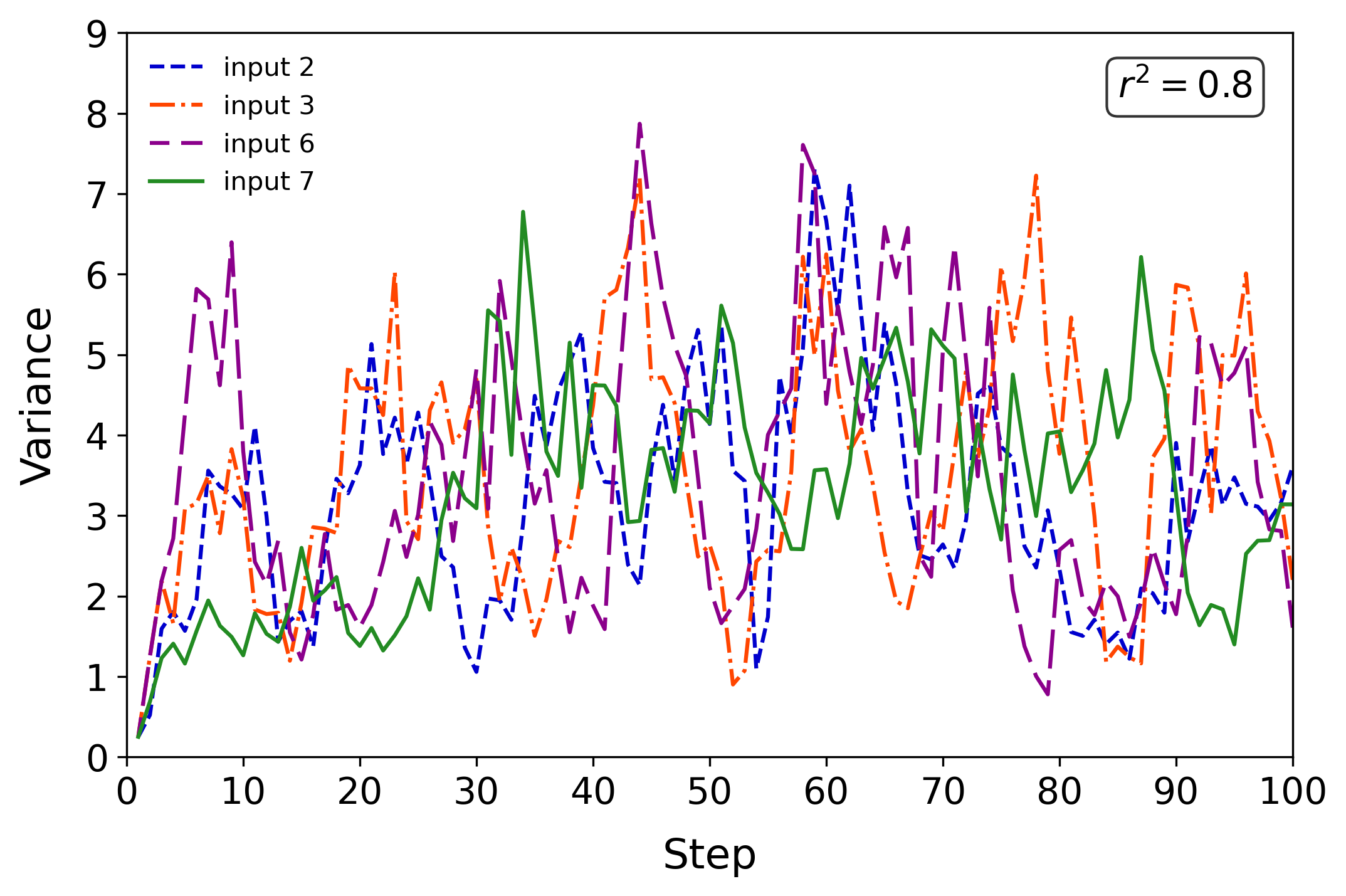}
    (d)
\end{minipage}
\caption{Simulated mean position (left column, panels (a) and (c)) and variance of the mean position (right column, panels (b) and (d)) as a function of the number of steps: comparison between the four selected input modes for leaking probability $r^2 = 0.2$ (panels (a) and (b)) and $r^2=0.8$ (panels (c) and (d)), corresponding to low- and high-leakage regime, respectively.}
\label{fig:Sim_100steps_cfr_Input}
\end{figure}

Due to the confinement, the walker propagation across the lattice follows oscillatory patterns, resulting from a sequence of total or partial reflections at the edges of the accessible region. This behavior is captured by the time evolution of both the mean position and its variance (Figure~\ref{fig:Sim_100steps_cfr_Input}\,(a)-(d)).
In the low-loss regime ($r^2=0.2$), the dynamics appears to be only weakly affected by the presence of the leaking boundary and, as shown in Figure~\ref{fig:Sim_100steps_cfr_Input}\,(a), the mean position depends on the injection site: when the walker enters the lattice from sites closer to one of the boundaries (input sites 2 and 7, cyan dashed and green solid lines), it propagates towards the opposite edge, whereas for more internal injection sites (3 and 6, orange dash-dotted and long-dashed pink lines) the mean position exhibits oscillations of smaller amplitude around the origin of the position axis. Nevertheless, the symmetry associated with the spatial coordinates of the input sites is largely preserved: opposite injection sites (2 and 7, or 3 and 6) share essentially the same temporal evolution, but with opposite phase.   
Likewise, the time evolution of the variance of the walker’s mean position exhibits an oscillatory behavior (Figure~\ref{fig:Sim_100steps_cfr_Input}\,(b)). For each pair of symmetric input sites, the variance follows the same qualitative temporal evolution, indicating that the spreading and partial refocusing of the wavefunction are similarly affected by the lattice boundaries when the leaking probability is low. In particular, when the walker is injected close to either the leaking or the reflecting edge (cyan dashed and green solid lines), the initial spreading remains limited and the variance stays below 3 for approximately the first 25 steps, before gradually increasing at longer times while preserving fluctuations of large amplitude and high frequency. On the other hand, for more internal injection sites the variance displays pronounced oscillations for the earliest steps, corresponding to alternative phases of spreading and relocalization of the wavefunction, occurring with a characteristic period of about 15 steps (orange dash-dotted and long-dashed pink lines).

By contrast, in the high-loss regime ($r^2=0.8$), the dynamics of both the mean position and its variance, although still retaining oscillatory features, departs from the regularities and phase relations identified in the low-loss case (Figure~\ref{fig:Sim_100steps_cfr_Input}\,(c),(d)). In particular, symmetric injection sites no longer display identical dynamics, and both the amplitude and frequency of the oscillations are modified, indicating a stronger impact of losses on the walker propagation.  
To further clarify the role of increasing losses on the system dynamics, it is helpful to compare the time evolution of the walker's mean position and variance for different leaking probabilities and injection sites. The full set of long-time numerical results is presented in Appendix~\ref{AppendixA_Results}.  

As expected, the strongest sensitivity to losses is observed for input site 2, which is the closest to the leaking boundary (Figure~\ref{fig:Sim_100steps_cfr_T}\,(a)-(b)). As long as the losses remain weak ($r^2=0.2$, cyan solid line), the dynamics closely follow that of the lossless case ($r^2=0$, gray dotted line), with few differences limited to their relative amplitudes up to approximately 40-50 steps. However, when the losses become strong ($r^2=0.8$, blue dashed line), deviations appear already at early times, with a faster propagation of the mean position (for $r^2=0.2$ the mean position reaches the first oscillation peak within the first 18 steps, whereas for $r^2=0.8$ it requires only 12 steps to reach the first maximum mean displacement) and an enhanced spreading of the wavefunction, as evidenced by the larger variance.  
For injection sites progressively farther from the leaking boundary (input sites 3, 6, and 7), the impact of losses on the walker's dynamics appears to be reduced (Figure~\ref{fig:Sim_100steps_cfr_T}\,(c)-(h)). In these configurations, the mean position and variance exhibit similar trends across different leaking regimes for a large number of time steps, with only minor differences in amplitude and phase. This indicates that, when the walker is injected sufficiently far from the lossy edge, the dynamics remain largely governed by coherent propagation within the lattice.  
In the following, we focus in particular on the initial $20$ time steps, for which a direct comparison with experimental observations was feasible.

\subsection{Measurements}
We experimentally implemented a quantum walk on a finite lattice featuring asymmetric boundary conditions—one reflective edge and one lossy (leaking) edge—using a universal photonic processor. The lattice was realized by employing 8 of the available 20 photonic modes. A leaking boundary was implemented at mode 1, while mode 8 acted as a perfectly reflecting boundary.

The quantum walker was injected at different initial modes within the lattice, either near the leaking boundary (waveguides 2 and 3) or near the reflecting boundary (waveguides 6 and 7). Two distinct leakage regimes were explored by tuning the reflection parameter to $r^2=0.2$ for weak leaking and $r^2=0.8$ for strong leaking. For each configuration, the walker was allowed to evolve for a number of steps ranging from 4 to 20. At each step, we measured the spatial probability distribution of the surviving wavefunction within the lattice. From these distributions, the mean position $\langle x\rangle$ and the variance $\sigma^2_n(x)$ of the walker according to Equations \ref{eq:variance}.

\begin{figure}[h]
\centering
\begin{minipage}{0.48\textwidth}
\centering
\includegraphics[width=\linewidth]{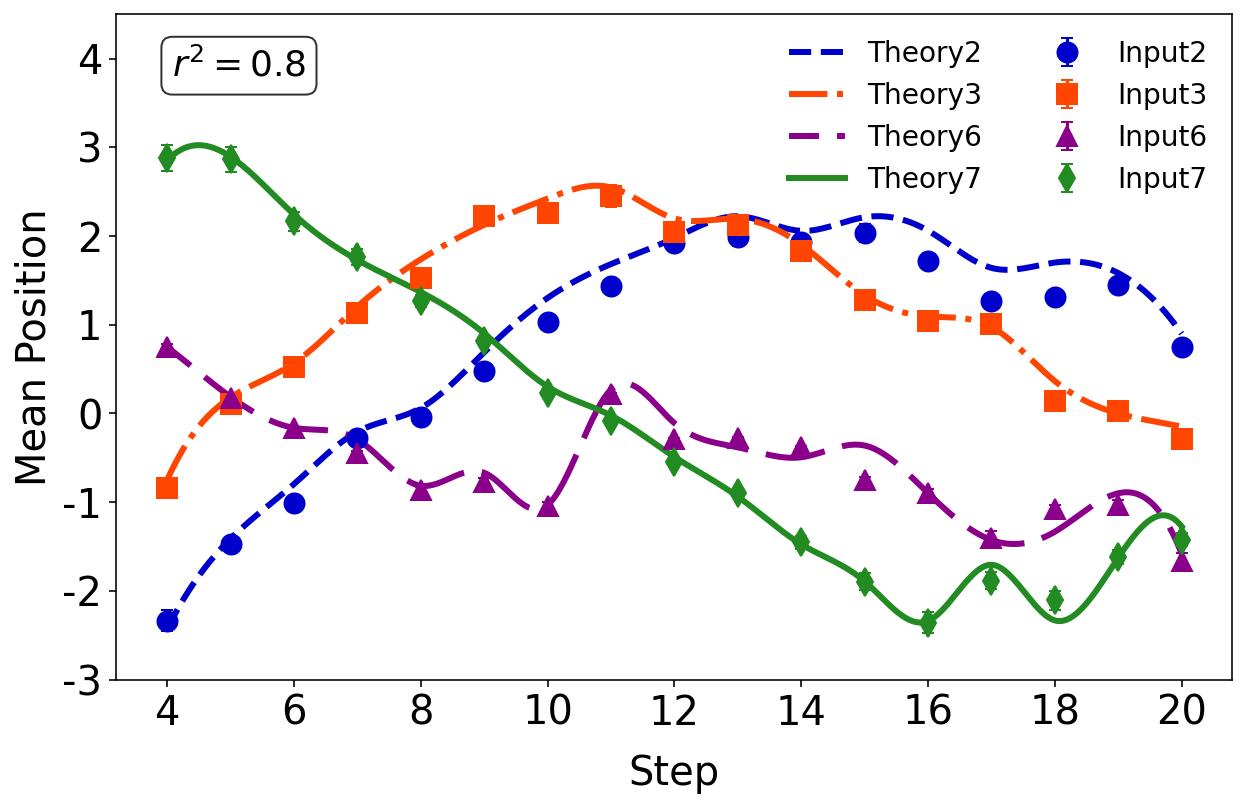}
\end{minipage}
\begin{minipage}{0.48\textwidth}
\includegraphics[width=\linewidth]{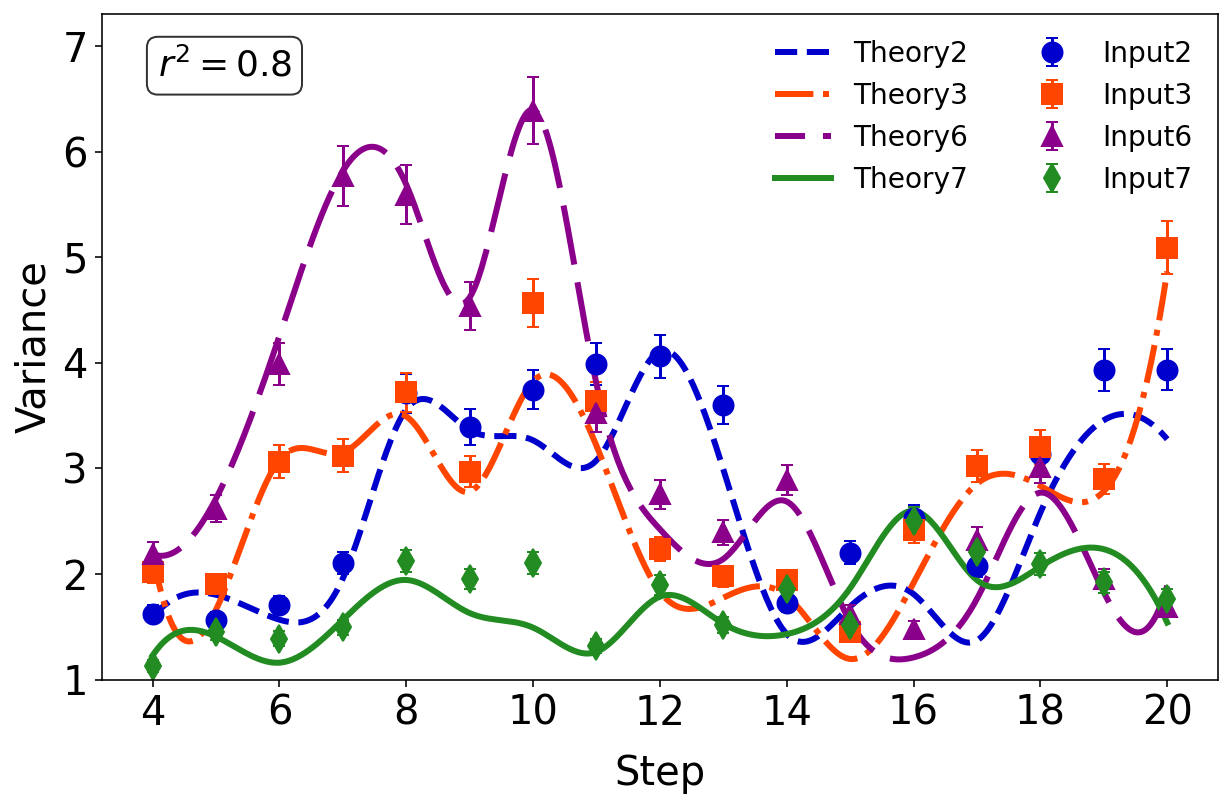}
\end{minipage}
\caption{Mean position (left) and variance of the mean position (right) as a function of the number of steps in the case of leaking probability $r^2=0.8$ corresponding to a strongly leaking boundary for different initial sites of the walk, namely site 2 corresponding to position $x=-2.5$ (blue circles), site 3 ($x=-1.5$ orange squares), site 6 ($x=1.5$, violet triangles) and site 7 ($x=2.5$, green diamonds). Symbols correspond to experimental data, lines to theoretical behaviors.}
\label{fig:R08}
\end{figure}

The mean position and variance for the strong-leakage regime are shown in Figure~\ref{fig:R08}. We observe good agreement between the experimental data (symbols) and the numerically simulated behavior (lines). The mean position reveals a net propagation of the walker across the lattice from its initial location toward the opposite boundary. When the walker is initialized near the leaking edge (input 2), it rapidly propagates toward the reflecting boundary, reaching an average displacement of approximately four lattice sites from the initial position after 15 steps (blue circles in Figure~\ref{fig:R08}, left panel). By contrast, when the walker is injected near the reflecting boundary (input 7), its propagation is initially slower; nevertheless, it attains a larger overall displacement, exceeding six lattice sites and reaching the position farthest from the initial mode after 16 steps (green diamonds).
The variance of the position reveals distinct spreading behaviors depending on the input site and the measurements (symbols) show a good agreement with the simulated curves (lines). However, here, when the walker is injected close to the reflecting boundary, the wavefunction exhibits limited spreading, as evidenced by the nearly flat variance curve for input 7 (green diamonds in Figure~ \ref{fig:R08}, right panel). For more central input sites, the wavefunction initially spreads and subsequently relocalizes (input 6, violet triangles), and in some cases spreads again (orange squares). When the input position is close to the leaking boundary, the wavefunction still displays spreading and relocalization dynamics; however, the amplitude of the spreading is reduced (blue circles).

\begin{figure}[t]
\centering
\begin{minipage}{0.48\textwidth}
\centering
\includegraphics[width=\linewidth]{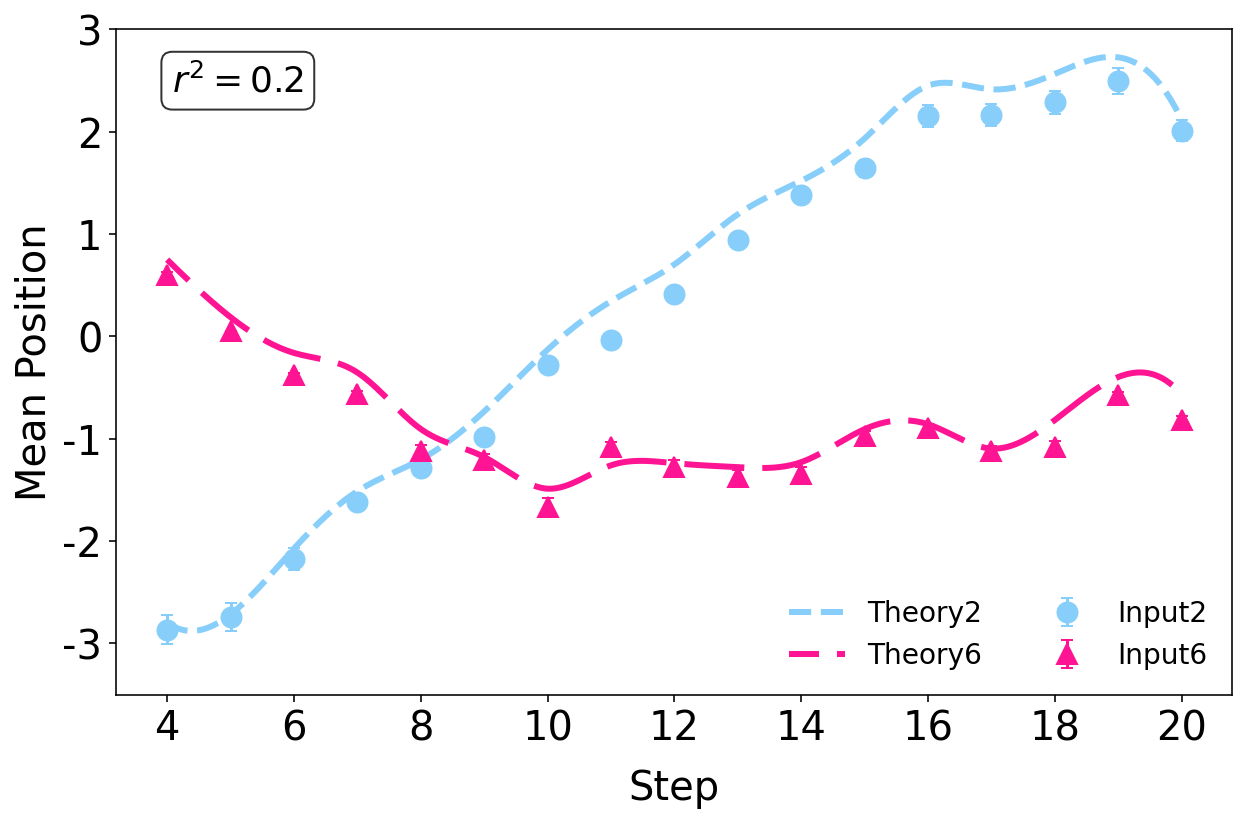}
\end{minipage}
\begin{minipage}{0.48\textwidth}
\includegraphics[width=\linewidth]{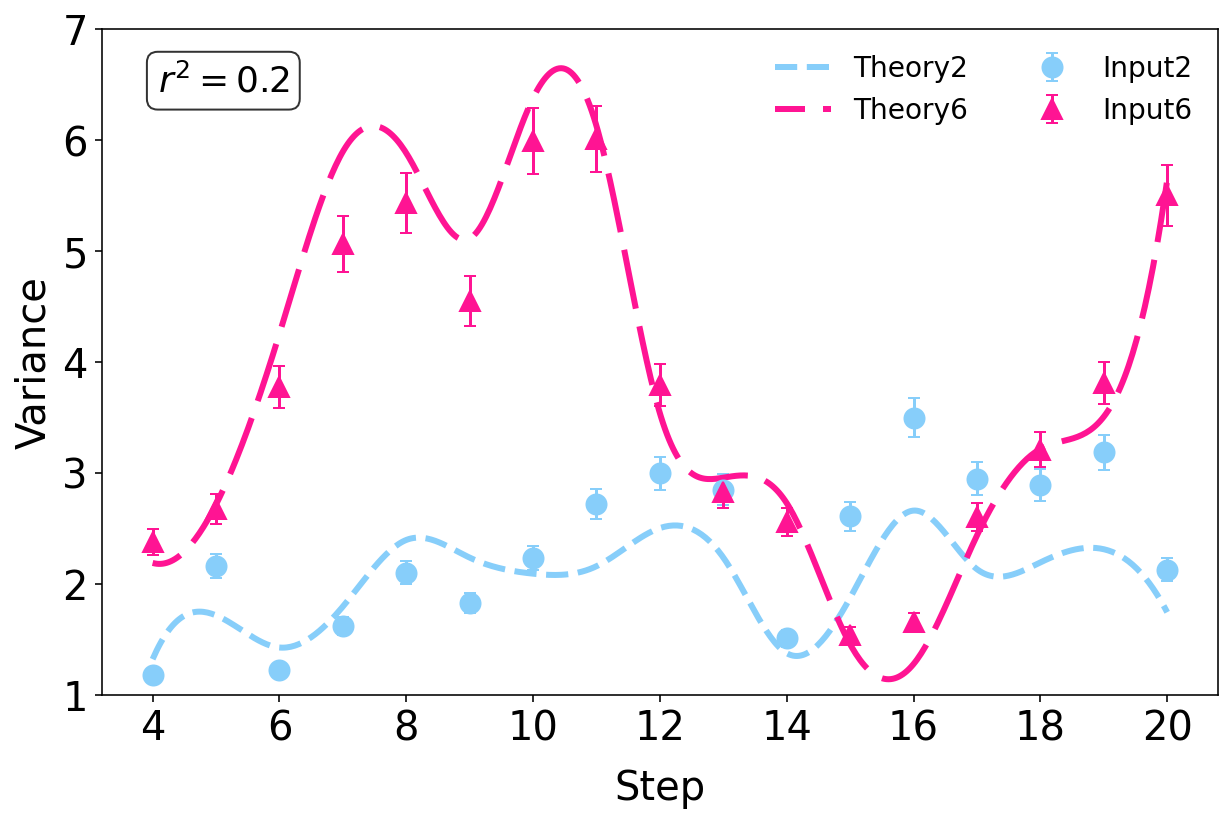}
\end{minipage}
\caption{Mean position (left) and variance of the mean position (right) as a function of the number of steps in the case of leaking probability $r^2=0.2$ corresponding to a weakly leaking boundary for different initial sites of the walk, namely site 2 corresponding to position $x=-2.5$ (cyan circles) and site 6 ($x=1.5$, pink triangles). Symbols correspond to experimental data, lines to theoretical behaviors.}
\label{fig:R02}
\end{figure}

We also measured the same quantities (mean position of the walker and its variance) in the case of a weak leaking boundary, by setting the parameters of our photonic processor such that the transmissivity of the beam splitters at the boundary is $r^2=0.2$. We measured the output probability distribution of the walker when injected in waveguide 2 and 6, close to the leaking and the reflecting boundaries respectively. Results are reported in Figure~\ref{fig:R02}.
Here we can observe a good agreement between experimental (symbols) and simulated (lines) data. It is interesting to note that if the walker initial position is close to the absorbing boundary, it moves toward the other edge more slowly than in the case of strong leaking, reaching the farthest position of its walk at step 19 (cyan circles in Figure~\ref{fig:R08} left panel). However its displacement is larger, reaching a value close to $\Delta x=6$ (compared to $\Delta x\approx 4$ in the previous configuration). For initial position corresponding to waveguide 6, the walker's mean position oscillates with a small displacement of around $\Delta x= 3$ (pink triangles). When looking at the variance of the mean position, a similar behavior to the configuration of strong leaking is observed: if the initial position is close to the reflecting boundary, the wavefunction spreads and relocalizes leading to an oscillating behavior of the variance (pink triangles in Figure~\ref{fig:R02} right panel), while a limited spreading almost without oscillations is observed if the walker enters the lattice close to the leaking boundary (cyan circles).

\begin{figure}[t]
\centering
\begin{minipage}{0.48\textwidth}
\centering
\includegraphics[width=\linewidth]{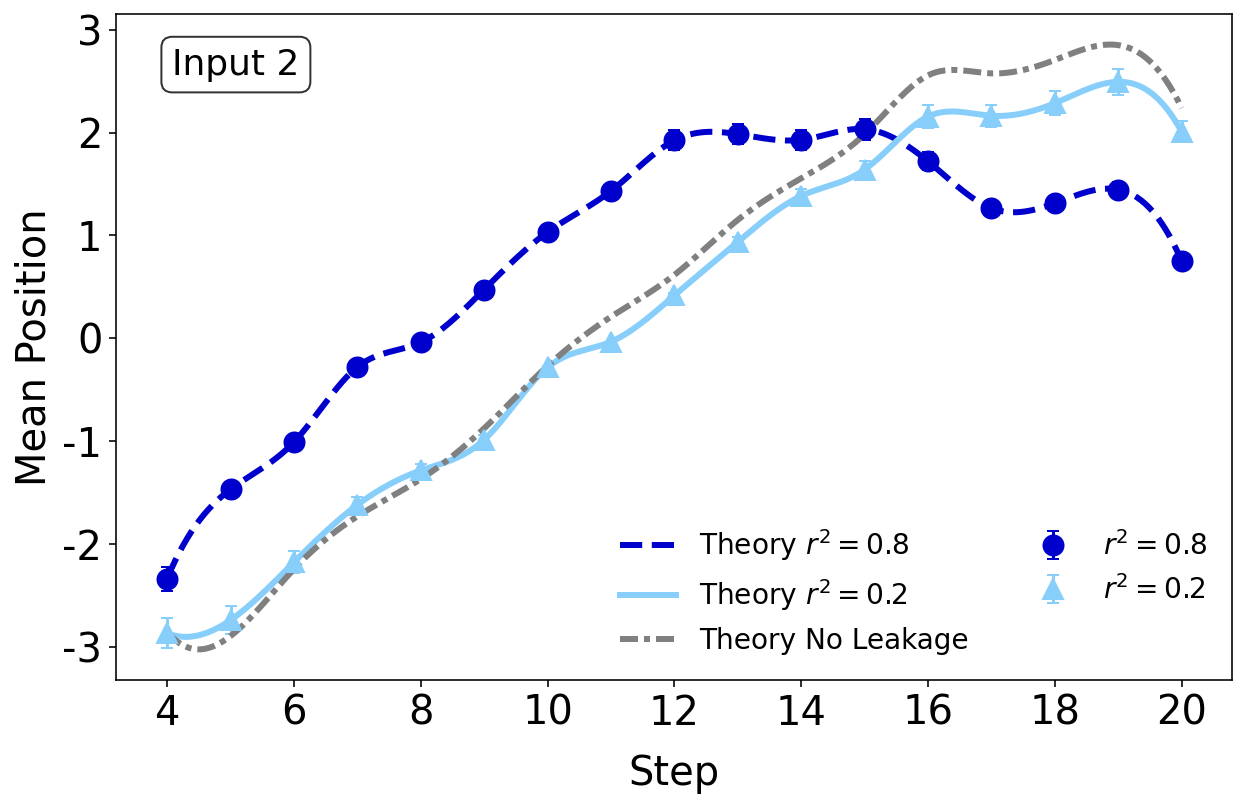}
\end{minipage}
\begin{minipage}{0.48\textwidth}
\includegraphics[width=\linewidth]{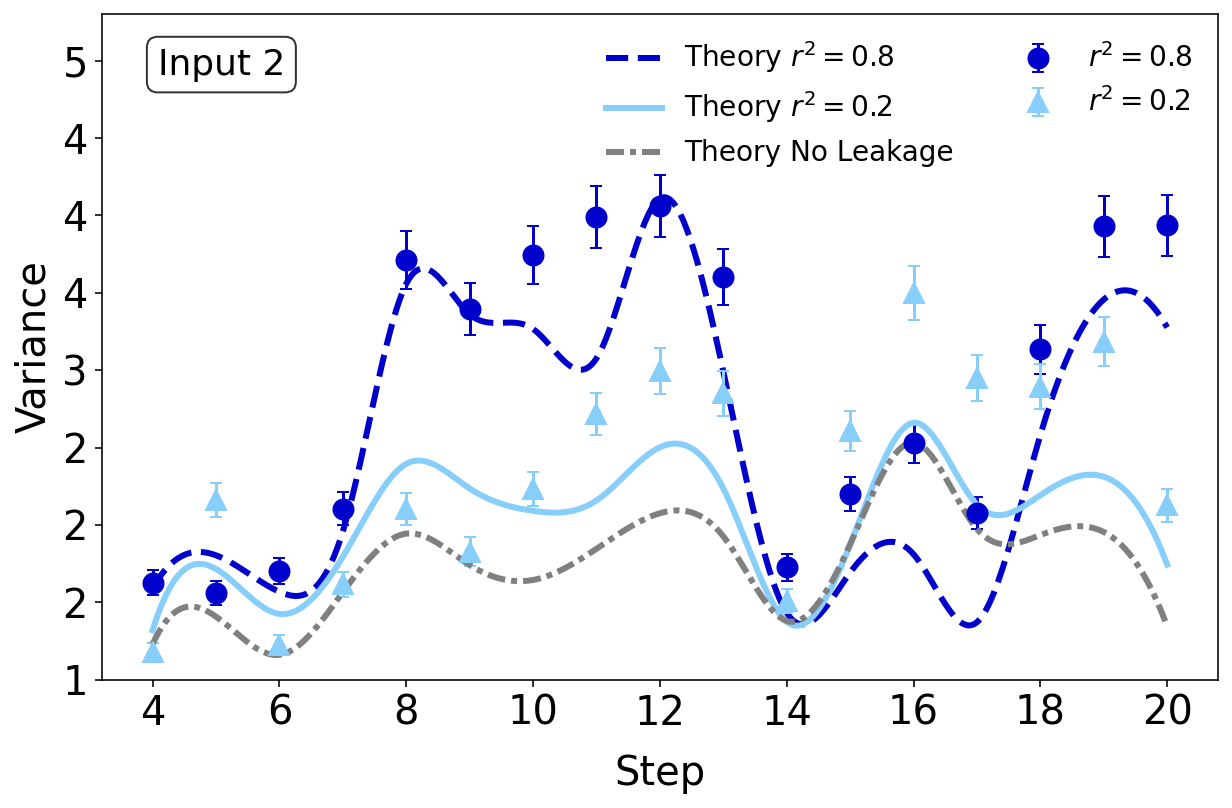}
\end{minipage}
\caption{Mean position (left) and variance of the mean position (right) as a function of the number of steps for a walker injected in mode 2. Blue circles correspond to strong leaking (transmissivity $r^2=0.8$), while cyan triangles to weak leaking ($r^2=0.2$). Symbols correspond to experimental data, lines to simulations. Gray dash-dotted lines show the behavior in absence of leaking.}
\label{fig:Input2}
\end{figure}

\begin{figure}[h]
\centering
\begin{minipage}{0.48\textwidth}
\centering
\includegraphics[width=\linewidth]{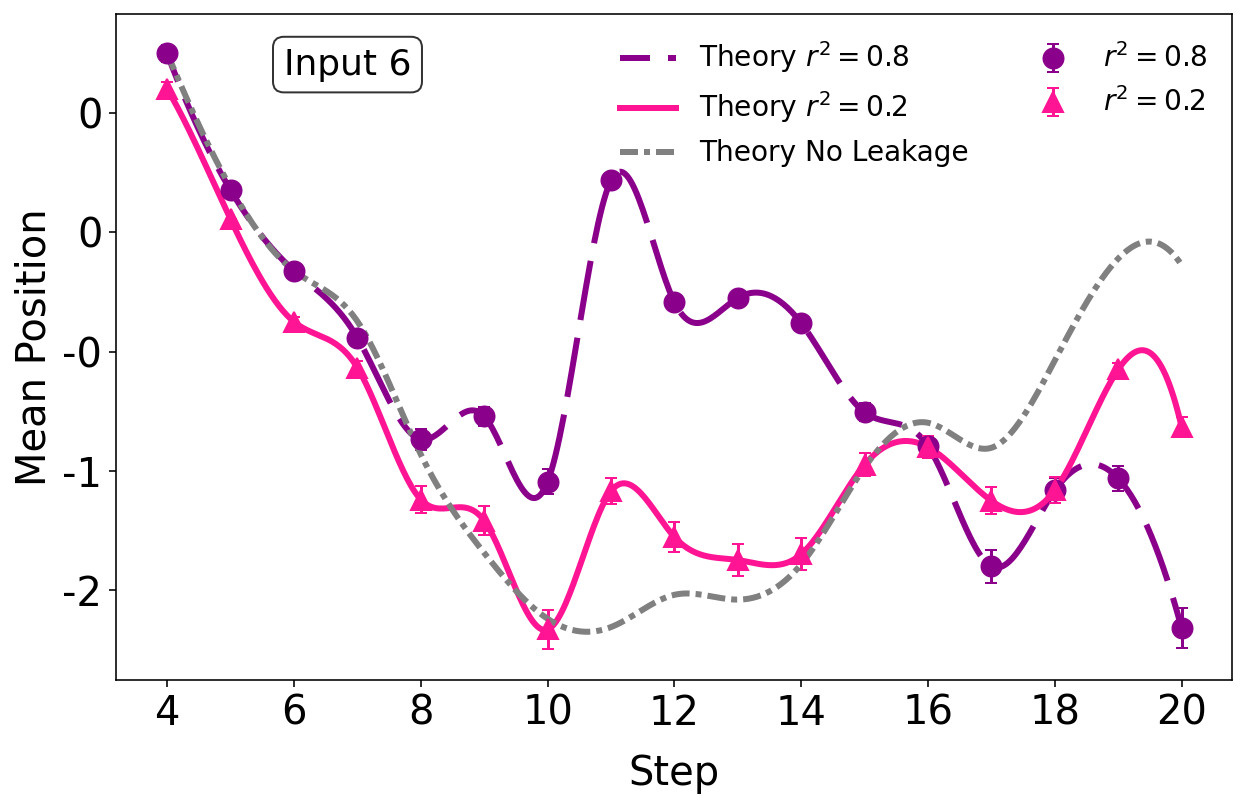}
\end{minipage}
\begin{minipage}{0.48\textwidth}
\includegraphics[width=\linewidth]{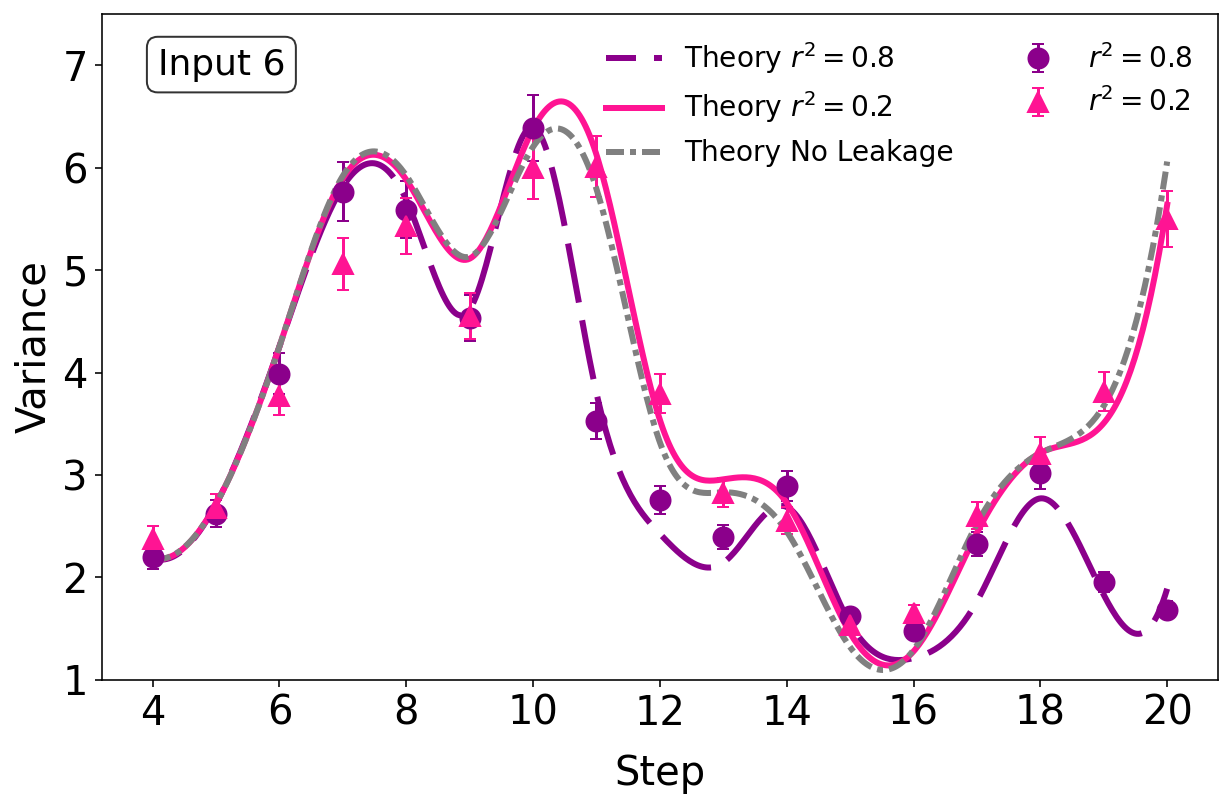}
\end{minipage}
\caption{Mean position (left) and variance of the mean position (right) as a function of the number of steps for a walker injected in site 6. Violet circles correspond to strong leaking (transmissivity $r^2=0.8$), while pink triangles to weak leaking ($r^2=0.2$). Symbols correspond to experimental data, lines to simulations. Gray dash-dotted lines show the behavior in absence of leaking.}
\label{fig:Input6}
\end{figure}

As a final comparison, we analyze the mean position and variance of the walker wavefunction for two different input sites, namely waveguides 2 and 6, in two different configurations of leakage probability. These behaviors are compared with numerical simulations of a quantum walk on a lattice without leakage, corresponding to fully reflective boundary conditions at both edges. The results are shown in Figures~\ref{fig:Input2} and \ref{fig:Input6} for inputs 2 and 6, respectively.
As predicted by the simulations discussed in the previous section, when the input site is close to the leaking edge, the weak-leakage regime (cyan triangles and solid line in Figure~\ref{fig:Input2}, left panel) closely reproduces the behavior observed in the absence of leakage (gray dash-dotted line). In contrast, at higher leakage probabilities (blue dots and dashed blue line), the dynamics differs significantly: the walker propagates more rapidly and reaches its maximum displacement at approximately step 13, although this displacement is reduced with respect to the other two cases $\Delta x\approx 5$ compared to $\Delta x\approx6$. This behavior is also reflected in the variance (Figure~\ref{fig:Input2}, right panel), where the strong-leakage regime exhibits oscillations with a larger amplitude (blue circles and dashed line) than those observed in the no-leakage or weak-leakage cases (gray dash-dotted line and cyan triangles with solid line, respectively).

Overall, when the walker is initialized near the leaking boundary, both the mean position and the variance are enhanced in the strong-leakage regime compared to the weak-leakage and no-leakage scenarios, as shown in Figure~\ref{fig:Input2}.
When the input site is far from the leaking boundary, a markedly different behavior is observed (Figure~\ref{fig:Input6}). During the first 8 steps of the walk, the dynamics are nearly identical across all three regimes. Subsequently, in the strong-leakage case (violet circles and dashed line), the mean position exhibits oscillations with a higher frequency compared to the other two regimes, whose behavior remains almost unchanged (Figure~\ref{fig:Input6}, left panel). By contrast, the variance displays very similar dynamics for all three regimes up to 18 steps (Figure~\ref{fig:Input6}, right panel), with only a slightly faster relocalization observed in the strong-leakage case between steps 8 and 13.

\section{Discussion}\label{Sec:Discussion}
The numerical results reveal that the DTQW dynamics is affected by boundary-induced losses in a position-dependent and nontrivial manner. While the presence of confinement leads to oscillatory propagation of the walker, the degree to which coherence and interference are preserved depends on both the leaking probability and the distance of the injection site from the lossy boundary. 

In the weak-leaking regime, the dynamics closely resembles that of an ideal, lossless QW. The symmetry between opposite injection sites is preserved in both the mean position and the variance, indicating that the partial leakage does not substantially disrupt long-range interference effects. The oscillatory behavior of the variance, characterized by alternating spreading and shrinking of the wavefunction, further confirms that the dynamics remain mainly coherent. In this regime, losses result only in a moderate attenuation of the amplitudes, without qualitatively altering the structure of the walk. 

As the leaking probability increases, deviations become progressively more pronounced. In the high-loss regime, the spoiling of the symmetry between opposite injection sites, together with changes in oscillation frequencies and amplitudes, suggests the onset of decoherence mechanisms induced by the strong openness of the system and the increased coupling to the environment.  

The comparison between different leaking regimes for the injection mode 2, which is the closest to the lossy boundary, highlights the strongest impact of these effects. While weak losses leave the dynamics largely unchanged with respect to the lossless case, strong leakage leads to faster propagation of the mean position as well as enhanced spreading of the wavefunction.
Additional features of the long-time dynamics, including the emergence of time windows characterized by enhanced or reduced variance and flattening of the mean displacement, are discussed in Appendix~\ref{AppendixA_Results_Discussion}.
Dynamical differences between the two limiting scenarios, i.e. absence of losses and a fully absorbing boundary, are particularly pronounced for the injection site closest to the leaking edge, confirming the key role played by boundary proximity in shaping both the early- and long-time dynamics. A detailed comparison between the lossless and fully absorbing boundary conditions is also reported in Appendix~\ref{AppendixA_Results_Discussion}. 

For injection modes farther from the leaking boundary, losses play a less significant role. The strong similarities between the dynamics observed in the weak- and high-loss regimes suggest that the wavefunction reaches the leaking boundary only with reduced probability and at later times, hence the overall evolution of both the mean position and the variance remains largely governed by coherent propagation. Remarkably, even the total-loss regime can preserve a dynamics closely resembling that of the ideal QW when the injection site is sufficiently distant from the leaking boundary, as pointed out by comparison between dotted and dashed curves in Figure~\ref{fig:Sim_100steps_cfr_T}\,(c)-(f). 

Overall, these findings are confirmed by experimental measurements performed on the photonic processor, which show good agreement with the simulated mean position and variance across different input sites and leakage regimes (Figures~\ref{fig:R08}, \ref{fig:R02}, \ref{fig:Input2}, \ref{fig:Input6}). Importantly, the experiments allow a closer inspection of the first 20 steps, highlighting subtle differences in the early-time propagation that are less visible in the long-time simulations.

The case of input site 7 deserves particular attention. Acting as a fully reflective barrier toward the interior of the lattice and being maximally distant from the leaking boundary, it would be expected to show the weakest sensitivity to losses. This is confirmed, but only at early times: up to about $15$ steps, both the mean position and the variance are nearly identical for all loss regimes, indicating that the wavefunction has not yet reached the opposite leaking edge (Figure~\ref{fig:Sim_100steps_cfr_T}\,(g),(h)). At longer times, however, a qualitative difference emerges. Weak losses lead to a pronounced flattening of the mean position and a drift towards positions closer to the leaking side, whereas stronger losses preserve oscillatory trends. A detailed analysis of the long-time behavior corresponding to this injection site can be found in Appendix~\ref{AppendixA_Results_Discussion}.

Taken together, the results discussed above indicate that losses in this system introduce a non-Hermitian leakage of the probability amplitude across the output modes, but do not necessarily induce a significant decoherence effect. From a general perspective, the resulting long-time dynamics exhibits a strong robustness in the presence of losses, especially in the low-leakage regime: the overall structure and the main qualitative features of the evolution patterns are only weakly affected by the introduction of a leaking boundary. This behavior stands in sharp contrast to scenarios in which dynamic noise is added to the system, where the wavefunction spreading and its propagation velocity can undergo substantial modifications \cite{Sansoni2025arxiv}. Conversely, when the leaking probability becomes high, a stronger coupling to the environment leads to a breakdown of dynamical regularities and phase relations still preserved at low losses, thereby mimicking some features typically associated with decoherence. 

Finally, it is worth noting that, in both the numerical model and its experimental implementation, we assumed a constant probability for a photon reaching the leaking boundary to escape the lattice at each time step. This choice allowed us to isolate the effect of spatially localized losses, avoiding to superimpose contributions from their time dependence. If the loss coefficients (i.e., the BS transmissivity at the boundary) were step-dependent, different propagation paths would be unevenly weighted, likely resulting in a partial suppression of the interference effects and potentially introducing temporal memory in the interference pattern. From an experimental perspective, assuming the leakage probability independent of the specific time step at which the walker reaches the boundary corresponds to an idealized scenario which was expected to be more controllable and suitable for a first investigation of the role played by a localized partially absorbing boundary. A more general and systematic study of the effects of time-dependent or randomly fluctuating losses, which would model more complex and realistic scenarios, is left for future works.  

\section{Conclusions}

Confined quantum walk with absorbing boundaries was investigated in detail. Within this framework, our results demonstrate that boundary leakage does not merely suppress QW dynamics, but can qualitatively reshape it in ways depending on both the loss strength and the injection geometry. From the experimental perspective, our findings show that recently developed commercial integrated photonic platforms provide a suitable architecture for the simulation of open quantum systems, enabling the introduction of controlled interactions between the quantum system and its environment with the consequent achievement of different levels of decoherence. The universal photonic quantum processor employed in the experiments enables the implementation of the desired Hamiltonians with controllable losses, demonstrating that leaking boundaries can be exploited as effective control parameters for engineered on-chip QWs. This, in turn, allows for the tuning of coherence, interference, and transport properties in integrated photonic platforms. The proposed approach offers a versatile framework that can be extended to a wide range of open quantum systems, paving the way to the realistic use of these quantum simulators to study non-trivial dynamics. Based on this work and the possible benefits arising from the exploitation of controllable decoherence, the capabilities of quantum computational methods based on the decoherence-enhancement could be pushed far beyond their current operational boundaries.

\noindent \textbf{Author contributions:} Conceptualization, E.S., L.S. and A.C.; methodology, all; software, E.S., L.S., J.P., J.B.,A.G. and A.C.; formal analysis, E.S. and L.S.; investigation, E.S., A.C. and L.S.; resources, all; data curation, E.S. and L.S.; writing---original draft preparation, E.S. and L.S.; writing---review and editing, all; visualization, E.S. and L.S.; supervision, A.C.; project administration, L.S.; funding acquisition, A.C.. All authors have read and agreed to the published version of the manuscript.

\noindent \textbf{Funding}: This research was funded by QuantERA II Programme supported by the EU H2020 research and innovation Programme under GA No 101017733, with funding Italian organization PNRR MUR project PE0000023-NQSTI (Spoke 6, CUP: H43C22000870001).

\noindent \textbf{Data availability}: The datasets used and analyzed during the current study are available from the corresponding author on reasonable request.

\noindent \textbf{Acknowledgments}: {The authors acknowledge Caterina Taballione from Quix Quantum, Marco Barbieri and Ilaria Gianani from Roma Tre University for fruitful and constructive conversations.}

\noindent \textbf{Conflicts of interest}: {The authors declare no conflicts of interest.}

\newpage
\appendix
\label{AppendixA_Results}
\section{Long-time numerical simulations}\label{AppendixA_Results_Results}

This appendix provides an extended analysis of the numerical simulations for the DTQW on a finite lattice in the presence of a leaking boundary, focusing on the long-time dynamics up to $100$ time steps. In particular, we report here on those dynamical features that were only briefly mentioned in Section~\ref{sec:Results_sim}, with the aim of supporting and clarify our physical interpretation.
These results complement the main text by providing a detailed comparison between different leakage regimes and injection sites.

\begin{figure}[ht!]
\centering
\begin{minipage}{0.48\textwidth}
    \centering
    \includegraphics[width=\linewidth]{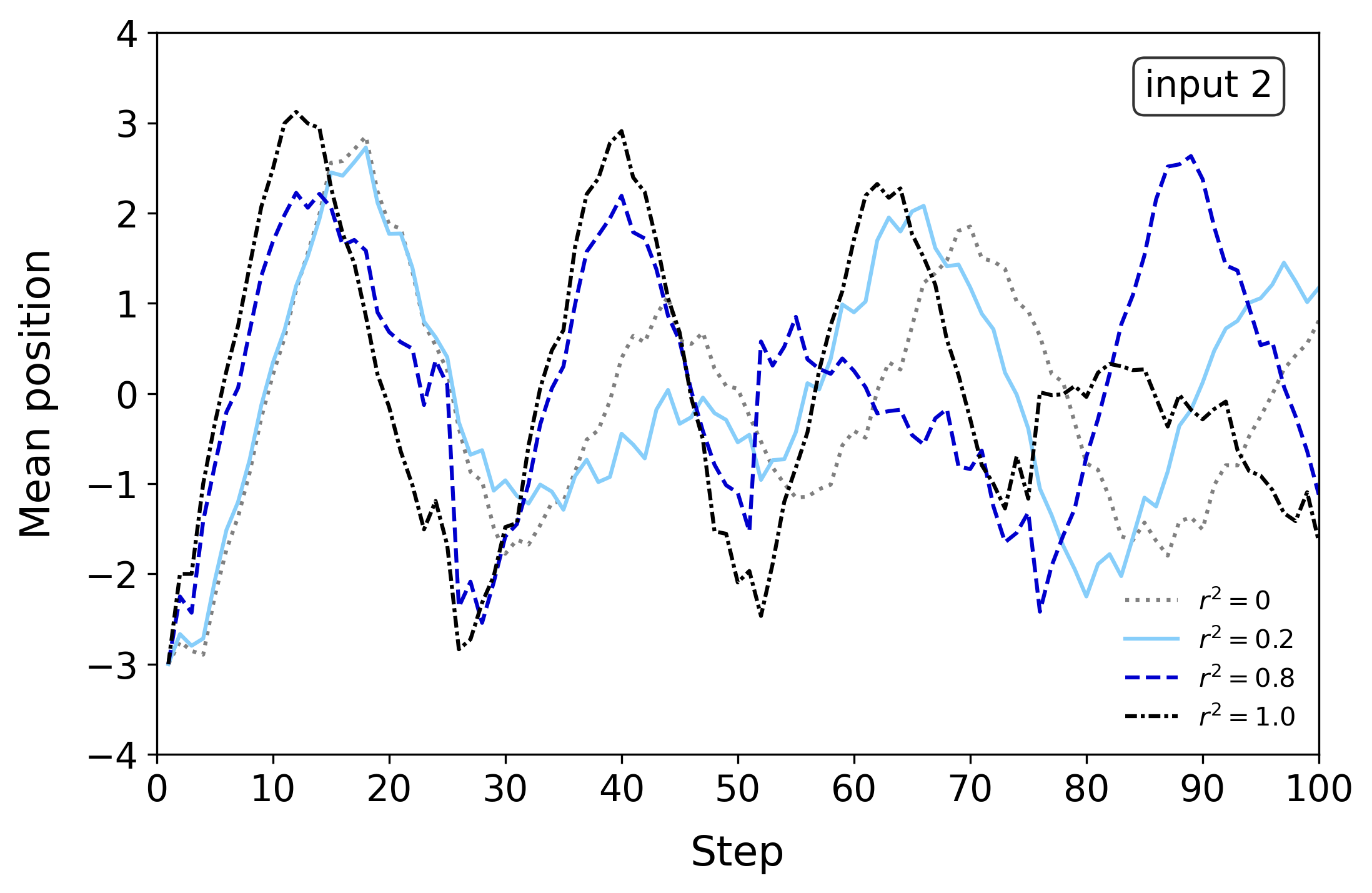}
    (a)
\end{minipage}\hfill
\begin{minipage}{0.48\textwidth}
    \centering
    \includegraphics[width=\linewidth]{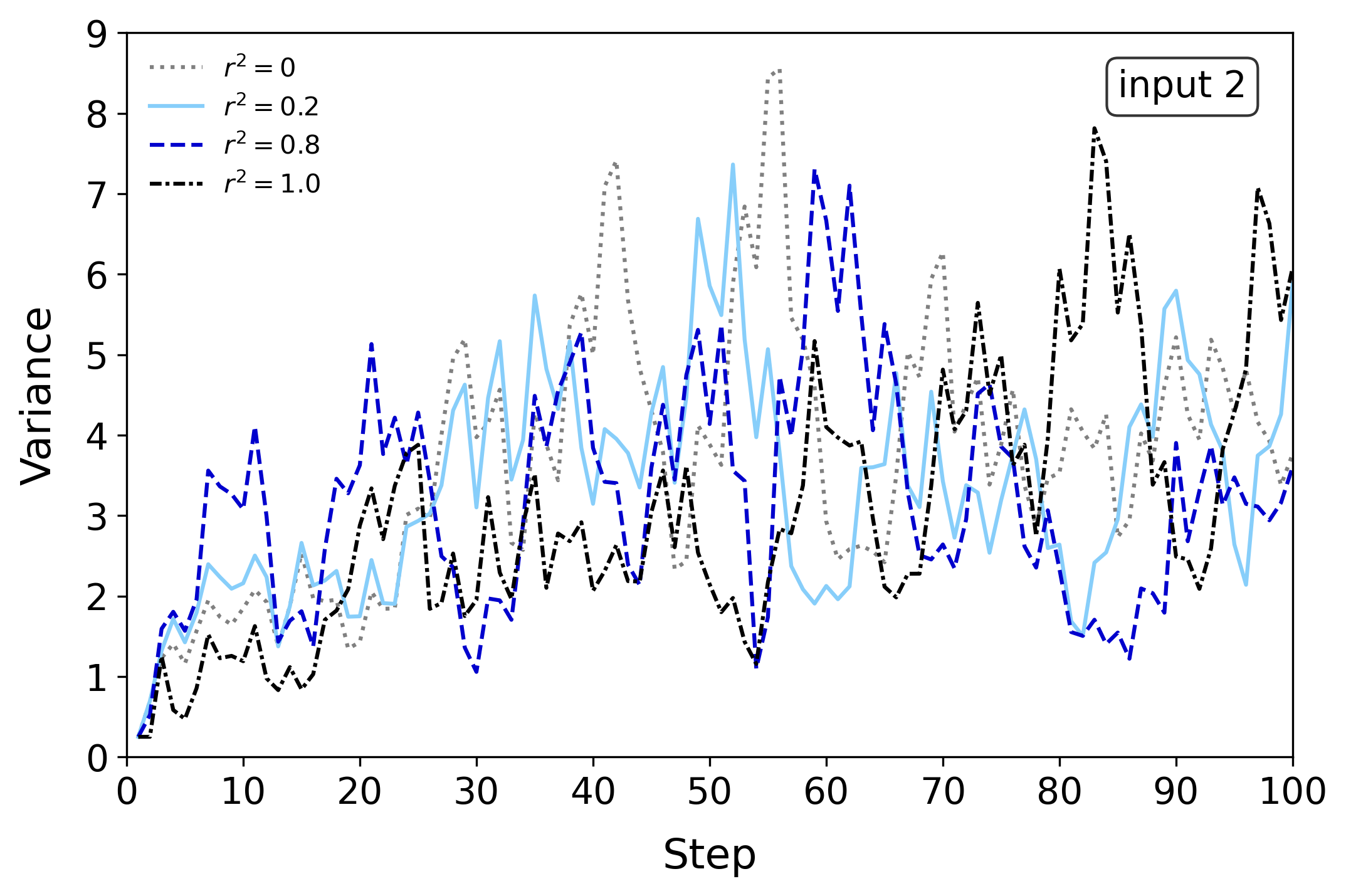}
    (b)
\end{minipage}
\vspace{0.3cm}
\begin{minipage}{0.48\textwidth}
    \centering
    \includegraphics[width=\linewidth]{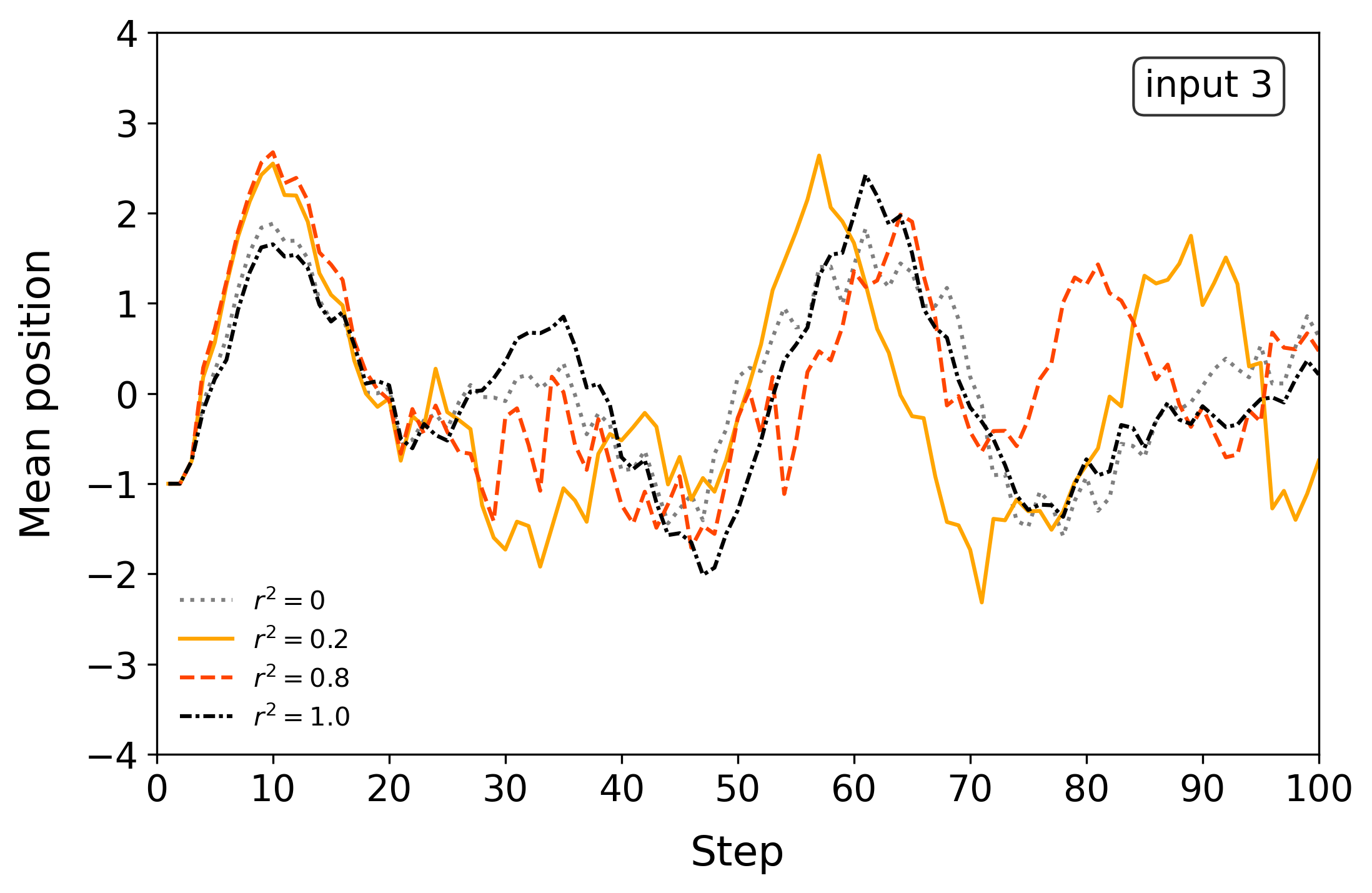}
    (c)
\end{minipage}\hfill
\begin{minipage}{0.48\textwidth}
    \centering
    \includegraphics[width=\linewidth]{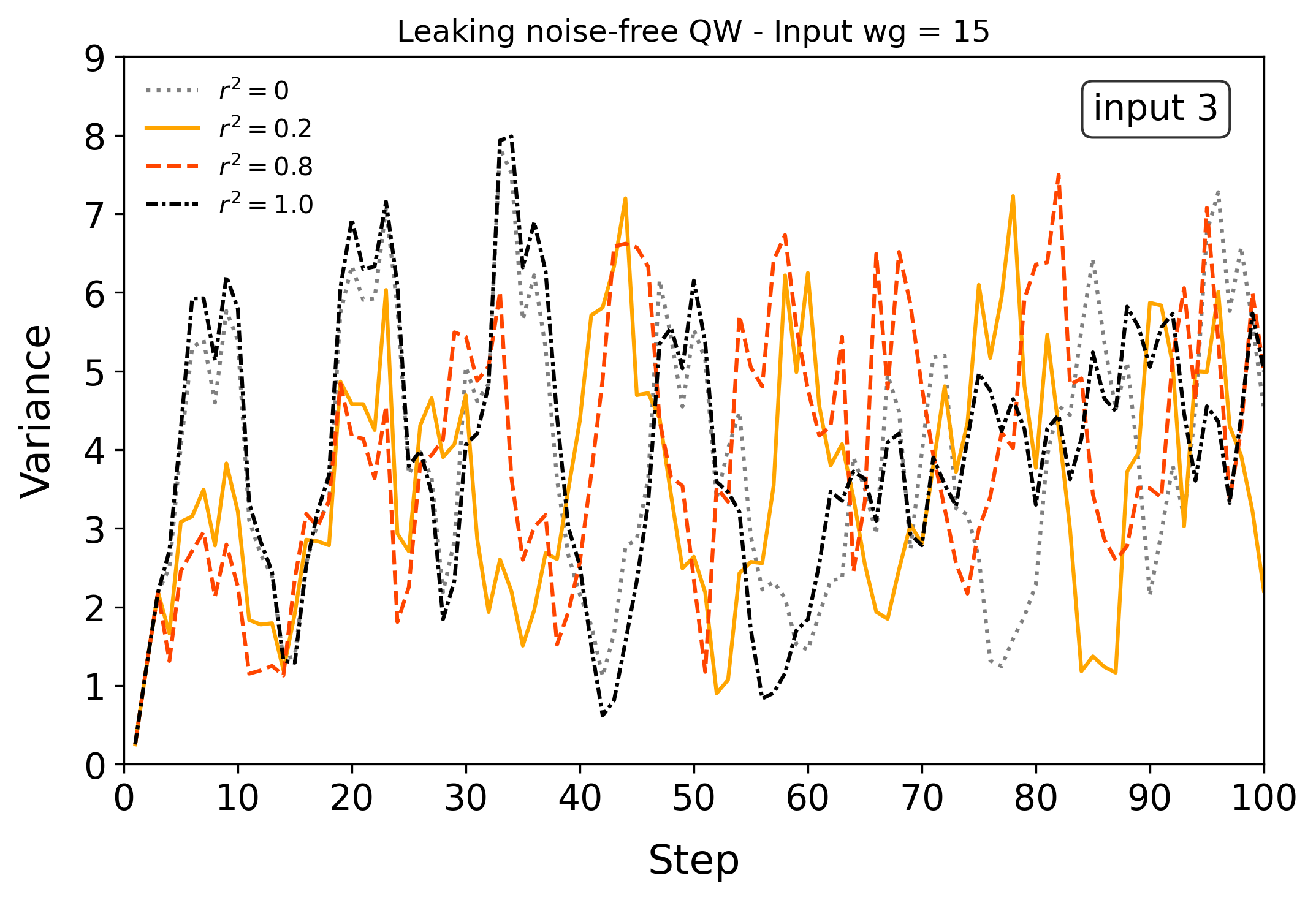}
    (d)
\end{minipage}

\begin{minipage}{0.48\textwidth}
    \centering
    \includegraphics[width=\linewidth]{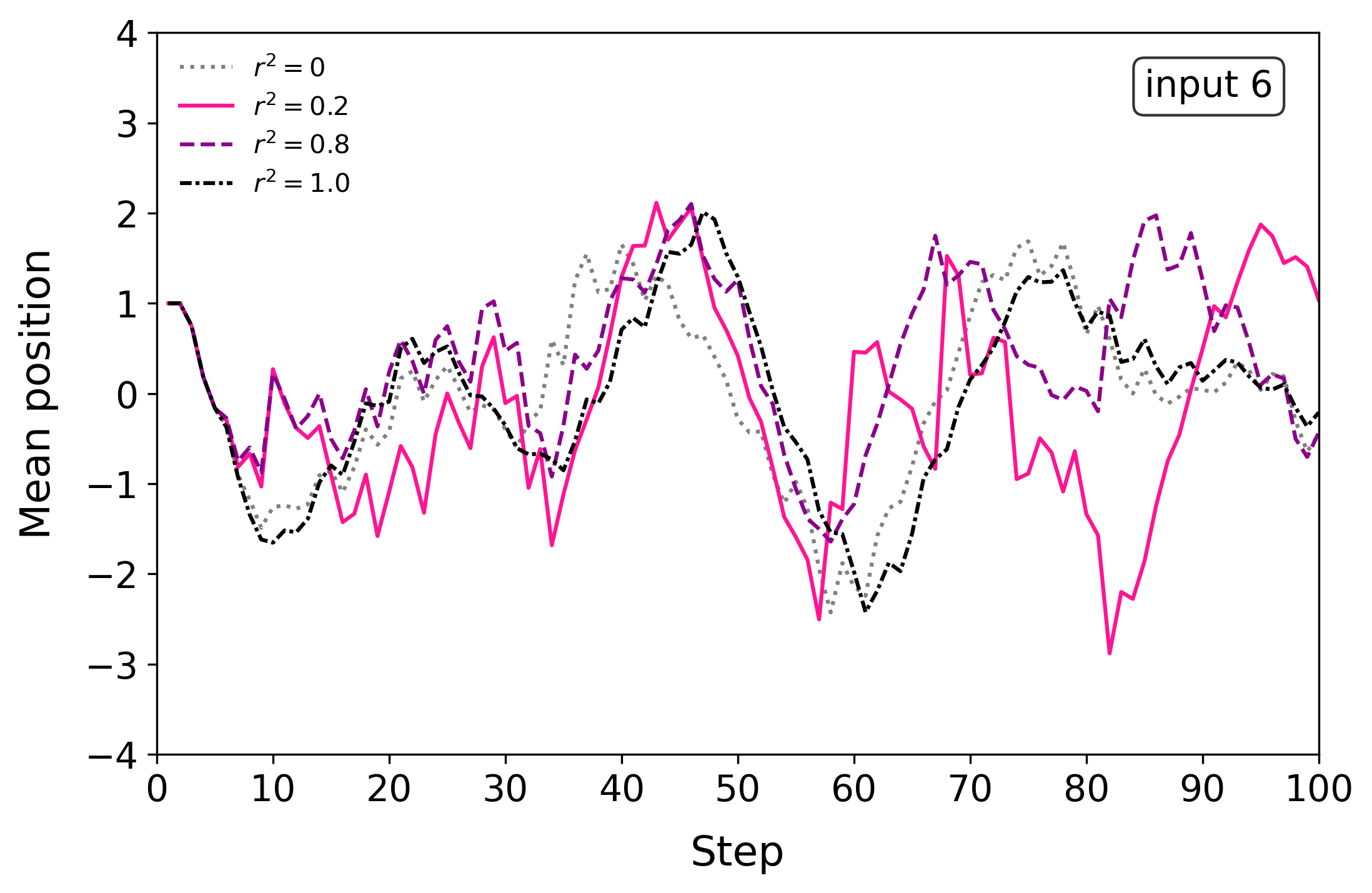}
    (e)
\end{minipage}\hfill
\begin{minipage}{0.48\textwidth}
    \centering
    \includegraphics[width=\linewidth]{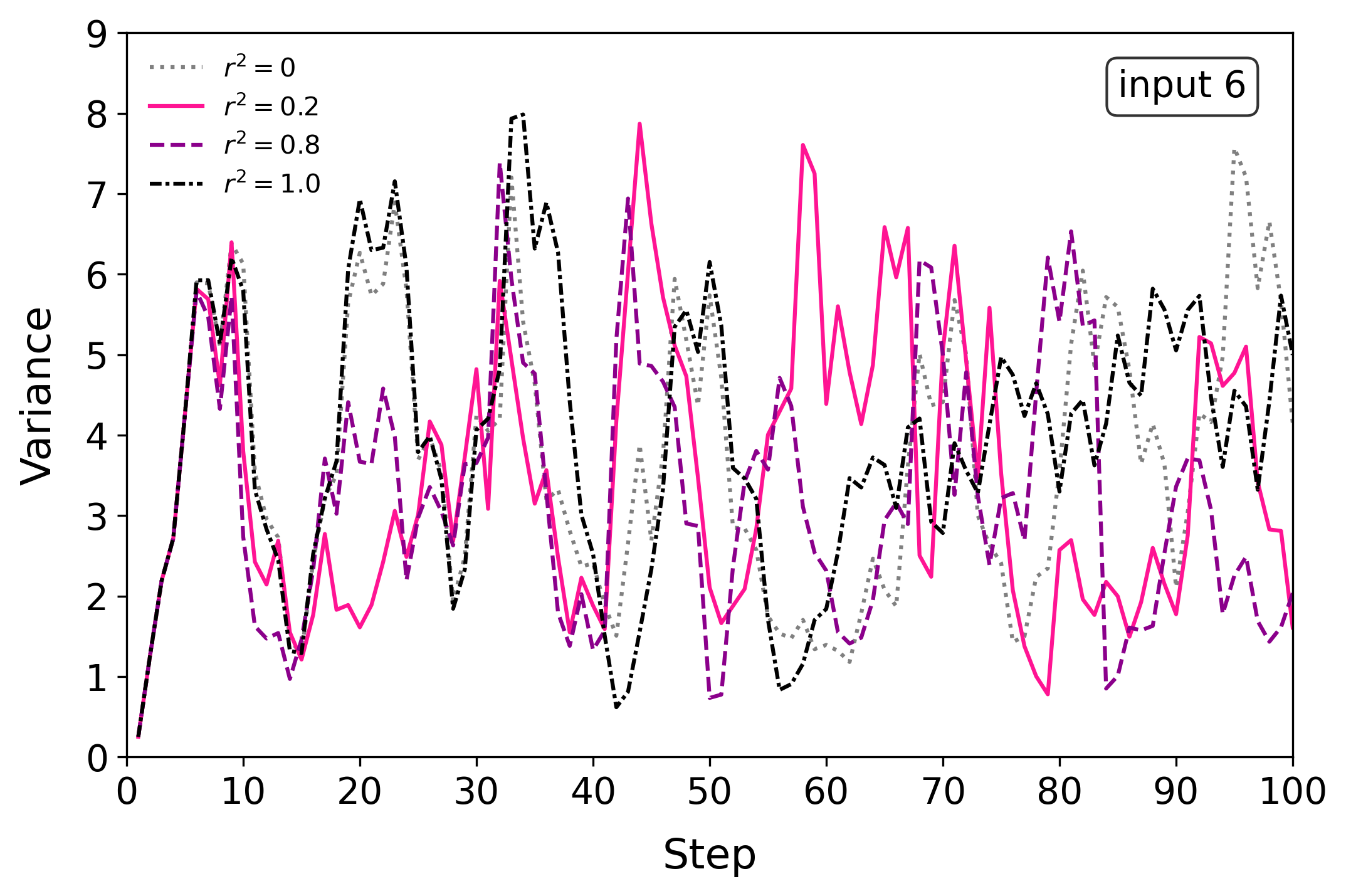}
    (f)
\end{minipage}

\begin{minipage}{0.48\textwidth}
    \centering
    \includegraphics[width=\linewidth]{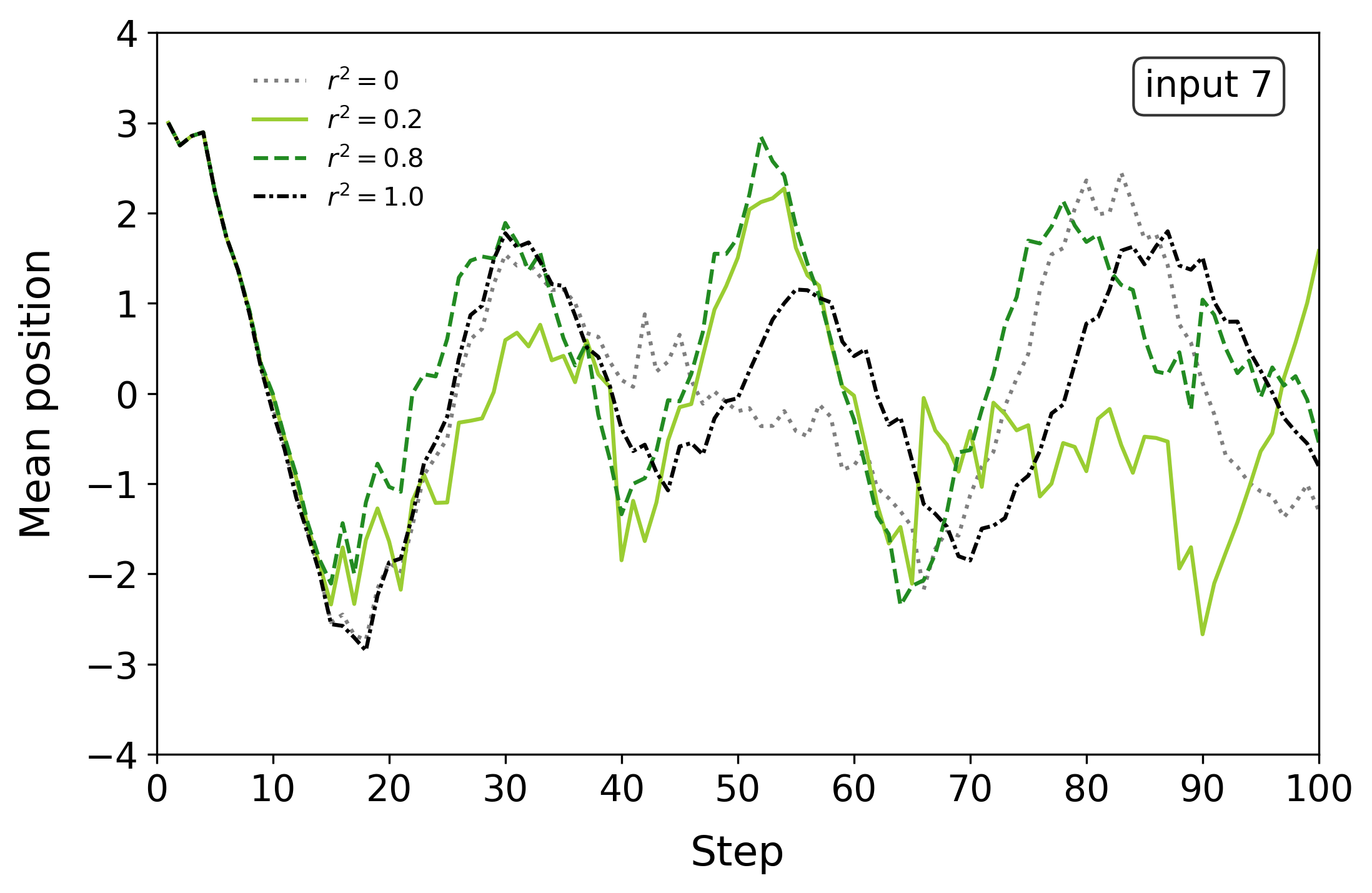}
    (g)
\end{minipage}\hfill
\begin{minipage}{0.48\textwidth}
    \centering
    \includegraphics[width=\linewidth]{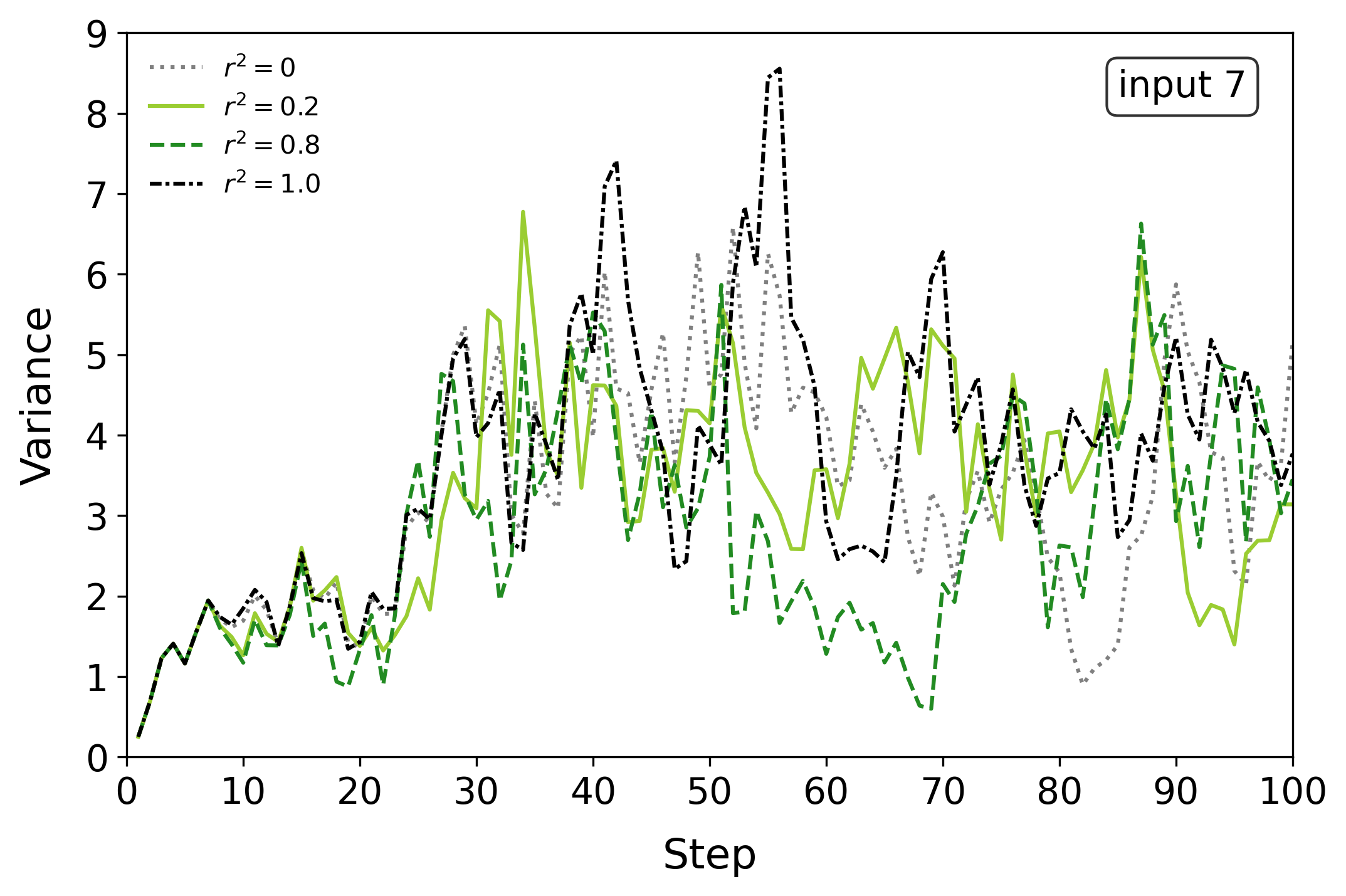}
    (h)
\end{minipage}
\caption{Simulated mean position (left column) and variance of the mean position (right column) as a function of the number of steps. Comparison between the low- (transmissivity $r^2=0.2$, colored solid lines) and high-leaking (transmissivity $r^2=0.8$, colored dashed lines) regimes for the four selected input modes: input 2 ((a)-(b)), input 3 ((c)-(d)), input 6 ((e)-(f)), and input 7 ((g)-(h)), where input 2 and input 7 correspond to the nearest and farthest injection sites from the leaking boundary. For reference, the lossless (transmissivity $r^2=0$) and the total loss (transmissivity $r^2=1.0$) cases are shown with gray dotted and black dash-dotted line, respectively.}
\label{fig:Sim_100steps_cfr_T}
\end{figure}

As reported in Section~\ref{sec:Results_sim}, the effects of losses strongly depend on the position of the injection site, as well as on the leakage strength. Figure~\ref{fig:Sim_100steps_cfr_T} therefore summarizes the time evolution of the walker's mean position and variance over $100$ steps for different injection sites and for the two representative leakage probabilities, namely $r^2=0.2$ (weak leakage) and $r^2=0.8$ (strong leakage). For reference, the lossless case ($r^2=0$) and the fully absorbing boundary ($r^2=1.0$) are also shown.

We first focus on the injection site closest to the leaking boundary (input 2, shown in Figure~\ref{fig:Sim_100steps_cfr_T}\,(a)-(b)), which in the main text was identified as the configuration most sensitive to losses.  
In the weak-leakage regime (cyan solid line), the dynamics closely resemble those of the lossless case (gray dotted line) over several tens of steps, with only moderate differences in amplitude. By contrast, strong leakage (blue dashed line) induces significant deviations already at early times, including a faster propagation of the mean position and an enhanced spreading of the wavefunction, as evidenced by the increased variance.
For injection sites progressively farther from the leaking boundary (input sites 3, 6, and 7 shown in Figure~\ref{fig:Sim_100steps_cfr_T}\,(c)-(d), (e)-(f), (g)-(h), respectively), the impact of losses is progressively reduced. In these configurations, both the mean position and the variance exhibit similar temporal trends across different leakage regimes over a large number of steps, indicating that the walker dynamics remain largely governed by coherent propagation within the lattice.

To further clarify the role of the leaking regimes discussed above, it is instructive to analyze the two limiting cases of fully reflective (transmissivity $r^2=0$) and fully absorbing (transmissivity $r^2=1.0$) boundaries on input 2, represented in Figure~\ref{fig:Sim_100steps_cfr_T} as gray dotted and black dashed-dotted lines, respectively. These limiting cases provide useful benchmarks for interpreting the intermediate leakage regimes discussed in the main text.

\section{Extended discussion on long-time dynamics}\label{AppendixA_Results_Discussion}

We provide here a more detailed interpretation of the long-time dynamical features briefly discussed in Section~\ref{Sec:Discussion}. 

Let's start our discussion from the injection site 2. Interestingly, if the walker is initialized closed to the leaking boundary, time windows emerge in which the mean position exhibits a less steep trend and stays close to the center of the lattice, while the variance reaches large values (e.g., between time steps 55 and 75, where the blue dashed curve in Figure~\ref{fig:Sim_100steps_cfr_T}\,(a) seems to lack a peak, while the blue dashed curve in Figure~\ref{fig:Sim_100steps_cfr_T}\,(b) displays its highest peak where the cyan solid line has a drop). This observation indicates that, whereas the absence of losses or their weak presence would lead to photon localization or bunching, strong leakage can suppress localization effects inducing a broader distribution of the probability amplitude across the waveguides while keeping the average position close to zero.
Injection site 2, due to its proximity to the leaking edge, results in the most pronounced dynamical differences between the two limiting scenarios, namely the absence of losses (transmissivity $r^2=0$) and the presence of a fully absorbing boundary (transmissivity $r^2=1.0$), with a leaking probability equal to 1. The lossless case is the ideal QW, where no photons escape the system: the walker's mean position exhibits the widest oscillations (gray dotted line in Figure~\ref{fig:Sim_100steps_cfr_T}\,(a)), indicating that the wavefunction is reflected between the lattice edges and interference preserves a high degree of coherence. This remark is reinforced by variance (gray dotted line in Figure~\ref{fig:Sim_100steps_cfr_T}\,(b)), that displays sharp peaks reaching values above 8. This implies that photons do not merely diffuse, but rather interfere, giving rise to pronounced interference lobes in the probability distribution. The sudden drops in variance point to transient re-focusing events, in which the wavefunction becomes more localized before spreading again. By contrast, in the total-loss regime (black dash-dotted line) the variance remains low for an extended time window (up to about the first 55 steps), indicating that most walkers are lost before contributing significantly to the spatial spreading. The appearance of a sharp peak in the variance at later times (beyond approximately 75 steps) can be attributed to the residual fraction of the surviving wavefunction reaching the lattice edges, which results in an increase in the computed variance even though the overall signal intensity is strongly reduced.

For the injection site maximally distant from the leaking boundary, i.e. input site 7, a counterintuitive behavior emerges at long time: although far from the absorbing edge leakage would be expected to have minimal influence over long times, weak losses cause a flattening of the mean position (between steps 65 and 90), as displayed by the green solid line in Figure~\ref{fig:Sim_100steps_cfr_T}\,(g), while stronger losses (green dashed line) preserve the oscillatory pattern. This effect can be ascribed to the fact that weak losses do not strongly suppress wavefunctions reaching the leaking edge, but instead partially attenuate and reinject them into the dynamics. This process breaks the interference symmetry and induces an effective directional bias. The corresponding variance dynamics supports this interpretation (Figure~\ref{fig:Sim_100steps_cfr_T}\,(h)). After an initial localization phase common to all regimes during the fist 20-25 steps, spreading occurs with reduce amplitude and broader peaks compared to those observed for more internal injection sites. Moreover, at longer times, variance of the high-leakage case (green dashed line) presents a marked drop around the 50th time step, revealing a transient relocalization of the wavefunction that is absent when losses are weaker (green solid line). This behavior mirrors that observed for the injection site 2, albeit with inverted roles between low- and high-loss cases, and underscores the nontrivial interplay between distance from the leaking boundary and loss-induced interference effects.

\clearpage
\printbibliography

@article{Aharonov1993,
  title = {Quantum random walks},
  volume = {48},
  ISSN = {1094-1622},
  url = {http://dx.doi.org/10.1103/PhysRevA.48.1687},
  DOI = {10.1103/physreva.48.1687},
  number = {2},
  journal = {Physical Review A},
  publisher = {American Physical Society (APS)},
  author = {Aharonov,  Y. and Davidovich,  L. and Zagury,  N.},
  year = {1993},
  pages = {1687–1690}
}

@article{Kempe2003,
  title = {Quantum random walks: An introductory overview},
  volume = {44},
  ISSN = {1366-5812},
  url = {http://dx.doi.org/10.1080/00107151031000110776},
  DOI = {10.1080/00107151031000110776},
  number = {4},
  journal = {Contemporary Physics},
  publisher = {Informa UK Limited},
  author = {Kempe,  J},
  year = {2003},
  pages = {307–327}
}

@article{VenegasAndraca2012,
  title = {Quantum walks: a comprehensive review},
  volume = {11},
  ISSN = {1573-1332},
  url = {http://dx.doi.org/10.1007/s11128-012-0432-5},
  DOI = {10.1007/s11128-012-0432-5},
  number = {5},
  journal = {Quantum Information Processing},
  publisher = {Springer Science and Business Media LLC},
  author = {Venegas-Andraca,  Salvador Elías},
  year = {2012},
  pages = {1015–1106}
}

@article{Childs2009,
  title = {Universal Computation by Quantum Walk},
  volume = {102},
  ISSN = {1079-7114},
  url = {http://dx.doi.org/10.1103/PhysRevLett.102.180501},
  DOI = {10.1103/physrevlett.102.180501},
  number = {18},
  journal = {Physical Review Letters},
  publisher = {American Physical Society (APS)},
  author = {Childs,  Andrew M.},
  year = {2009}
}

@article{AMBAINIS2003,
  title = {Quantum walks and their algorithmic applications},
  volume = {01},
  ISSN = {1793-6918},
  url = {http://dx.doi.org/10.1142/S0219749903000383},
  DOI = {10.1142/s0219749903000383},
  number = {04},
  journal = {International Journal of Quantum Information},
  publisher = {World Scientific Pub Co Pte Lt},
  author = {Ambainis,  Andris},
  year = {2003},
  pages = {507–518}
}

@article{Shenvi2003,
  title = {Quantum random-walk search algorithm},
  volume = {67},
  ISSN = {1094-1622},
  url = {http://dx.doi.org/10.1103/PhysRevA.67.052307},
  DOI = {10.1103/physreva.67.052307},
  number = {5},
  journal = {Physical Review A},
  publisher = {American Physical Society (APS)},
  author = {Shenvi,  Neil and Kempe,  Julia and Whaley,  K. Birgitta},
  year = {2003}
}

@article{AspuruGuzik2012,
  title = {Photonic quantum simulators},
  volume = {8},
  ISSN = {1745-2481},
  url = {http://dx.doi.org/10.1038/nphys2253},
  DOI = {10.1038/nphys2253},
  number = {4},
  journal = {Nature Physics},
  publisher = {Springer Science and Business Media LLC},
  author = {Aspuru-Guzik,  Alán and Walther,  Philip},
  year = {2012},
  pages = {285–291}
}

@article{Kitagawa2010,
  title = {Exploring topological phases with quantum walks},
  volume = {82},
  ISSN = {1094-1622},
  url = {http://dx.doi.org/10.1103/PhysRevA.82.033429},
  DOI = {10.1103/physreva.82.033429},
  number = {3},
  journal = {Physical Review A},
  publisher = {American Physical Society (APS)},
  author = {Kitagawa,  Takuya and Rudner,  Mark S. and Berg,  Erez and Demler,  Eugene},
  year = {2010},
}

@article{Broome2010,
  title = {Discrete Single-Photon Quantum Walks with Tunable Decoherence},
  volume = {104},
  ISSN = {1079-7114},
  url = {http://dx.doi.org/10.1103/PhysRevLett.104.153602},
  DOI = {10.1103/physrevlett.104.153602},
  number = {15},
  journal = {Physical Review Letters},
  publisher = {American Physical Society (APS)},
  author = {Broome,  M. A. and Fedrizzi,  A. and Lanyon,  B. P. and Kassal,  I. and Aspuru-Guzik,  A. and White,  A. G.},
  year = {2010}
}

@article{Peruzzo2010,
  title = {Quantum Walks of Correlated Photons},
  volume = {329},
  ISSN = {1095-9203},
  url = {http://dx.doi.org/10.1126/science.1193515},
  DOI = {10.1126/science.1193515},
  number = {5998},
  journal = {Science},
  publisher = {American Association for the Advancement of Science (AAAS)},
  author = {Peruzzo,  Alberto and Lobino,  Mirko and Matthews,  Jonathan C. F. and Matsuda,  Nobuyuki and Politi,  Alberto and Poulios,  Konstantinos and Zhou,  Xiao-Qi and Lahini,  Yoav and Ismail,  Nur and W\"{o}rhoff,  Kerstin and Bromberg,  Yaron and Silberberg,  Yaron and Thompson,  Mark G. and OBrien,  Jeremy L.},
  year = {2010},
  pages = {1500–1503}
}

@article{Sansoni2012,
  title = {Two-Particle Bosonic-Fermionic Quantum Walk via Integrated Photonics},
  volume = {108},
  ISSN = {1079-7114},
  url = {http://dx.doi.org/10.1103/PhysRevLett.108.010502},
  DOI = {10.1103/physrevlett.108.010502},
  number = {1},
  journal = {Physical Review Letters},
  publisher = {American Physical Society (APS)},
  author = {Sansoni,  Linda and Sciarrino,  Fabio and Vallone,  Giuseppe and Mataloni,  Paolo and Crespi,  Andrea and Ramponi,  Roberta and Osellame,  Roberto},
  year = {2012}
}

@article{Tang2018,
  title = {Experimental two-dimensional quantum walk on a photonic chip},
  volume = {4},
  ISSN = {2375-2548},
  url = {http://dx.doi.org/10.1126/sciadv.aat3174},
  DOI = {10.1126/sciadv.aat3174},
  number = {5},
  journal = {Science Advances},
  publisher = {American Association for the Advancement of Science (AAAS)},
  author = {Tang,  Hao and Lin,  Xiao-Feng and Feng,  Zhen and Chen,  Jing-Yuan and Gao,  Jun and Sun,  Ke and Wang,  Chao-Yue and Lai,  Peng-Cheng and Xu,  Xiao-Yun and Wang,  Yao and Qiao,  Lu-Feng and Yang,  Ai-Lin and Jin,  Xian-Min},
  year = {2018}
}

@article{Zhou2024,
  title = {Multi-particle quantum walks on 3D integrated photonic chip},
  volume = {13},
  ISSN = {2047-7538},
  url = {http://dx.doi.org/10.1038/s41377-024-01627-7},
  DOI = {10.1038/s41377-024-01627-7},
  number = {1},
  journal = {Light: Science \& Applications},
  publisher = {Springer Science and Business Media LLC},
  author = {Zhou,  Wen-Hao and Wang,  Xiao-Wei and Ren,  Ruo-Jing and Fu,  Yu-Xuan and Chang,  Yi-Jun and Xu,  Xiao-Yun and Tang,  Hao and Jin,  Xian-Min},
  year = {2024}
}

@article{Schreiber2010,
  title = {Photons Walking the Line: A Quantum Walk with Adjustable Coin Operations},
  volume = {104},
  ISSN = {1079-7114},
  url = {http://dx.doi.org/10.1103/PhysRevLett.104.050502},
  DOI = {10.1103/physrevlett.104.050502},
  number = {5},
  journal = {Physical Review Letters},
  publisher = {American Physical Society (APS)},
  author = {Schreiber,  A. and Cassemiro,  K. N. and Potoček,  V. and Gábris,  A. and Mosley,  P. J. and Andersson,  E. and Jex,  I. and Silberhorn,  Ch.},
  year = {2010}
}

@article{Schreiber2011,
  title = {Decoherence and Disorder in Quantum Walks: From Ballistic Spread to Localization},
  volume = {106},
  ISSN = {1079-7114},
  url = {http://dx.doi.org/10.1103/PhysRevLett.106.180403},
  DOI = {10.1103/physrevlett.106.180403},
  number = {18},
  journal = {Physical Review Letters},
  publisher = {American Physical Society (APS)},
  author = {Schreiber,  A. and Cassemiro,  K. N. and Potoček,  V. and Gábris,  A. and Jex,  I. and Silberhorn,  Ch.},
  year = {2011}
}

@article{Schreiber2012,
  title = {A 2D Quantum Walk Simulation of Two-Particle Dynamics},
  volume = {336},
  ISSN = {1095-9203},
  url = {http://dx.doi.org/10.1126/science.1218448},
  DOI = {10.1126/science.1218448},
  number = {6077},
  journal = {Science},
  publisher = {American Association for the Advancement of Science (AAAS)},
  author = {Schreiber,  Andreas and Gábris,  Aurél and Rohde,  Peter P. and Laiho,  Kaisa and Štefaňák,  Martin and Potoček,  Václav and Hamilton,  Craig and Jex,  Igor and Silberhorn,  Christine},
  year = {2012},
  pages = {55–58}
}

@article{Crespi2013,
  title = {Anderson localization of entangled photons in an integrated quantum walk},
  volume = {7},
  ISSN = {1749-4893},
  url = {http://dx.doi.org/10.1038/nphoton.2013.26},
  DOI = {10.1038/nphoton.2013.26},
  number = {4},
  journal = {Nature Photonics},
  publisher = {Springer Science and Business Media LLC},
  author = {Crespi,  Andrea and Osellame,  Roberto and Ramponi,  Roberta and Giovannetti,  Vittorio and Fazio,  Rosario and Sansoni,  Linda and De Nicola,  Francesco and Sciarrino,  Fabio and Mataloni,  Paolo},
  year = {2013},
  pages = {322–328}
}

@article{DeNicola2014,
  title = {Quantum simulation of bosonic-fermionic noninteracting particles in disordered systems via a quantum walk},
  volume = {89},
  ISSN = {1094-1622},
  url = {http://dx.doi.org/10.1103/PhysRevA.89.032322},
  DOI = {10.1103/physreva.89.032322},
  number = {3},
  journal = {Physical Review A},
  publisher = {American Physical Society (APS)},
  author = {De Nicola,  Francesco and Sansoni,  Linda and Crespi,  Andrea and Ramponi,  Roberta and Osellame,  Roberto and Giovannetti,  Vittorio and Fazio,  Rosario and Mataloni,  Paolo and Sciarrino,  Fabio},
  year = {2014}
}

@article{Dadras2018,
  title = {Quantum Walk in Momentum Space with a Bose-Einstein Condensate},
  volume = {121},
  ISSN = {1079-7114},
  url = {http://dx.doi.org/10.1103/PhysRevLett.121.070402},
  DOI = {10.1103/physrevlett.121.070402},
  number = {7},
  journal = {Physical Review Letters},
  publisher = {American Physical Society (APS)},
  author = {Dadras,  Siamak and Gresch,  Alexander and Groiseau,  Caspar and Wimberger,  Sandro and Summy,  Gil S.},
  year = {2018}
}

@article{Dadras2019,
  title = {Experimental realization of a momentum-space quantum walk},
  volume = {99},
  ISSN = {2469-9934},
  url = {http://dx.doi.org/10.1103/PhysRevA.99.043617},
  DOI = {10.1103/physreva.99.043617},
  number = {4},
  journal = {Physical Review A},
  publisher = {American Physical Society (APS)},
  author = {Dadras,  Siamak and Gresch,  Alexander and Groiseau,  Caspar and Wimberger,  Sandro and Summy,  Gil S.},
  year = {2019}
}

@article{Clark2021,
  title = {Quantum to classical walk transitions tuned by spontaneous emissions},
  volume = {3},
  ISSN = {2643-1564},
  url = {http://dx.doi.org/10.1103/PhysRevResearch.3.043062},
  DOI = {10.1103/physrevresearch.3.043062},
  number = {4},
  journal = {Physical Review Research},
  publisher = {American Physical Society (APS)},
  author = {Clark,  J. H. and Groiseau,  C. and Shaw,  Z. N. and Dadras,  S. and Binegar,  C. and Wimberger,  S. and Summy,  G. S. and Liu,  Y.},
  year = {2021}
}

@article{Gong2021,
  title = {Quantum walks on a programmable two-dimensional 62-qubit superconducting processor},
  volume = {372},
  ISSN = {1095-9203},
  url = {http://dx.doi.org/10.1126/science.abg7812},
  DOI = {10.1126/science.abg7812},
  number = {6545},
  journal = {Science},
  publisher = {American Association for the Advancement of Science (AAAS)},
  author = {Gong,  Ming and Wang,  Shiyu and Zha,  Chen and Chen,  Ming-Cheng and Huang,  He-Liang and Wu,  Yulin and Zhu,  Qingling and Zhao,  Youwei and Li,  Shaowei and Guo,  Shaojun and Qian,  Haoran and Ye,  Yangsen and Chen,  Fusheng and Ying,  Chong and Yu,  Jiale and Fan,  Daojin and Wu,  Dachao and Su,  Hong and Deng,  Hui and Rong,  Hao and Zhang,  Kaili and Cao,  Sirui and Lin,  Jin and Xu,  Yu and Sun,  Lihua and Guo,  Cheng and Li,  Na and Liang,  Futian and Bastidas,  V. M. and Nemoto,  Kae and Munro,  W. J. and Huo,  Yong-Heng and Lu,  Chao-Yang and Peng,  Cheng-Zhi and Zhu,  Xiaobo and Pan,  Jian-Wei},
  year = {2021},
  pages = {948–952}
}

@article{Wang2019,
  title = {Integrated photonic quantum technologies},
  volume = {14},
  ISSN = {1749-4893},
  url = {http://dx.doi.org/10.1038/s41566-019-0532-1},
  DOI = {10.1038/s41566-019-0532-1},
  number = {5},
  journal = {Nature Photonics},
  publisher = {Springer Science and Business Media LLC},
  author = {Wang,  Jianwei and Sciarrino,  Fabio and Laing,  Anthony and Thompson,  Mark G.},
  year = {2019},
  pages = {273–284}
}

@inproceedings{Aaronson2011,
  series = {STOC’11},
  title = {The computational complexity of linear optics},
  url = {http://dx.doi.org/10.1145/1993636.1993682},
  DOI = {10.1145/1993636.1993682},
  booktitle = {Proceedings of the forty-third annual ACM symposium on Theory of computing},
  publisher = {ACM},
  author = {Aaronson,  Scott and Arkhipov,  Alex},
  year = {2011},
  pages = {333–342},
  collection = {STOC’11}
}

@article{Lund2017,
  title = {Quantum sampling problems,  BosonSampling and quantum supremacy},
  volume = {3},
  ISSN = {2056-6387},
  url = {http://dx.doi.org/10.1038/s41534-017-0018-2},
  DOI = {10.1038/s41534-017-0018-2},
  number = {1},
  journal = {npj Quantum Information},
  publisher = {Springer Science and Business Media LLC},
  author = {Lund,  A. P. and Bremner,  Michael J. and Ralph,  T. C.},
  year = {2017}
}

@article{Hamilton2017,
  title = {Gaussian Boson Sampling},
  volume = {119},
  ISSN = {1079-7114},
  url = {http://dx.doi.org/10.1103/PhysRevLett.119.170501},
  DOI = {10.1103/physrevlett.119.170501},
  number = {17},
  journal = {Physical Review Letters},
  publisher = {American Physical Society (APS)},
  author = {Hamilton,  Craig S. and Kruse,  Regina and Sansoni,  Linda and Barkhofen,  Sonja and Silberhorn,  Christine and Jex,  Igor},
  year = {2017}
}

@article{Tillmann2013,
  title = {Experimental boson sampling},
  volume = {7},
  ISSN = {1749-4893},
  url = {http://dx.doi.org/10.1038/nphoton.2013.102},
  DOI = {10.1038/nphoton.2013.102},
  number = {7},
  journal = {Nature Photonics},
  publisher = {Springer Science and Business Media LLC},
  author = {Tillmann,  Max and Dakić,  Borivoje and Heilmann,  René and Nolte,  Stefan and Szameit,  Alexander and Walther,  Philip},
  year = {2013},
  pages = {540–544}
}

@article{Spagnolo2014,
  title = {Experimental validation of photonic boson sampling},
  volume = {8},
  ISSN = {1749-4893},
  url = {http://dx.doi.org/10.1038/nphoton.2014.135},
  DOI = {10.1038/nphoton.2014.135},
  number = {8},
  journal = {Nature Photonics},
  publisher = {Springer Science and Business Media LLC},
  author = {Spagnolo,  Nicolò and Vitelli,  Chiara and Bentivegna,  Marco and Brod,  Daniel J. and Crespi,  Andrea and Flamini,  Fulvio and Giacomini,  Sandro and Milani,  Giorgio and Ramponi,  Roberta and Mataloni,  Paolo and Osellame,  Roberto and Galvão,  Ernesto F. and Sciarrino,  Fabio},
  year = {2014},
  pages = {615–620}
}

@article{Crespi2013bis,
  title = {Integrated multimode interferometers with arbitrary designs for photonic boson sampling},
  volume = {7},
  ISSN = {1749-4893},
  url = {http://dx.doi.org/10.1038/nphoton.2013.112},
  DOI = {10.1038/nphoton.2013.112},
  number = {7},
  journal = {Nature Photonics},
  publisher = {Springer Science and Business Media LLC},
  author = {Crespi,  Andrea and Osellame,  Roberto and Ramponi,  Roberta and Brod,  Daniel J. and Galvão,  Ernesto F. and Spagnolo,  Nicolò and Vitelli,  Chiara and Maiorino,  Enrico and Mataloni,  Paolo and Sciarrino,  Fabio},
  year = {2013},
  pages = {545–549}
}

@misc{Anguita2025arx,
  doi = {10.48550/ARXIV.2509.25404},
  url = {https://arxiv.org/abs/2509.25404},
  author = {Anguita,  Malaquias Correa and Roelink,  Teun and Marzban,  Sara and Briels,  Wim and Filippi,  Claudia and Renema,  Jelmer},
  title = {Experimental demonstration of boson sampling as a hardware accelerator for monte carlo integration},
  publisher = {arXiv},
  year = {2025},
  copyright = {Creative Commons Attribution 4.0 International}
}

@article{Zhong2019,
  title = {Experimental Gaussian Boson sampling},
  volume = {64},
  ISSN = {2095-9273},
  url = {http://dx.doi.org/10.1016/j.scib.2019.04.007},
  DOI = {10.1016/j.scib.2019.04.007},
  number = {8},
  journal = {Science Bulletin},
  publisher = {Elsevier BV},
  author = {Zhong,  Han-Sen and Peng,  Li-Chao and Li,  Yuan and Hu,  Yi and Li,  Wei and Qin,  Jian and Wu,  Dian and Zhang,  Weijun and Li,  Hao and Zhang,  Lu and Wang,  Zhen and You,  Lixing and Jiang,  Xiao and Li,  Li and Liu,  Nai-Le and Dowling,  Jonathan P. and Lu,  Chao-Yang and Pan,  Jian-Wei},
  year = {2019},
  pages = {511–515}
}

@article{Hoch2022,
  title = {Reconfigurable continuously-coupled 3D photonic circuit for Boson Sampling experiments},
  volume = {8},
  ISSN = {2056-6387},
  url = {http://dx.doi.org/10.1038/s41534-022-00568-6},
  DOI = {10.1038/s41534-022-00568-6},
  number = {1},
  journal = {npj Quantum Information},
  publisher = {Springer Science and Business Media LLC},
  author = {Hoch,  Francesco and Piacentini,  Simone and Giordani,  Taira and Tian,  Zhen-Nan and Iuliano,  Mariagrazia and Esposito,  Chiara and Camillini,  Anita and Carvacho,  Gonzalo and Ceccarelli,  Francesco and Spagnolo,  Nicolò and Crespi,  Andrea and Sciarrino,  Fabio and Osellame,  Roberto},
  year = {2022}
}

@misc{Sansoni2025arxiv,
      title={Noisy dynamics of confined quantum walks on a chip}, 
      author={L. Sansoni and E. Stefanutti and C. Benedetti and I. Gianani and C. Taballione and A. Toor and L. Herrera and M. Pistilli and S. Santoro and M. Barbieri and A. Chiuri},
      year={2025},
      eprint={2511.19125},
      archivePrefix={arXiv},
      primaryClass={quant-ph},
      url={https://arxiv.org/abs/2511.19125}, 
}

@article{Taballione2023,
  title = {20-Mode Universal Quantum Photonic Processor},
  volume = {7},
  ISSN = {2521-327X},
  url = {http://dx.doi.org/10.22331/q-2023-08-01-1071},
  DOI = {10.22331/q-2023-08-01-1071},
  journal = {Quantum},
  publisher = {Verein zur Forderung des Open Access Publizierens in den Quantenwissenschaften},
  author = {Taballione,  Caterina and Anguita,  Malaquias Correa and de Goede,  Michiel and Venderbosch,  Pim and Kassenberg,  Ben and Snijders,  Henk and Kannan,  Narasimhan and Vleeshouwers,  Ward L. and Smith,  Devin and Epping,  J\"{o}rn P. and van der Meer,  Reinier and Pinkse,  Pepijn W. H. and van den Vlekkert,  Hans and Renema,  Jelmer J.},
  year = {2023},
  pages = {1071}
}

@article{Taballione2021,
  title = {A universal fully reconfigurable 12-mode quantum photonic processor},
  volume = {1},
  ISSN = {2633-4356},
  url = {http://dx.doi.org/10.1088/2633-4356/ac168c},
  DOI = {10.1088/2633-4356/ac168c},
  number = {3},
  journal = {Materials for Quantum Technology},
  publisher = {IOP Publishing},
  author = {Taballione,  Caterina and van der Meer,  Reinier and Snijders,  Henk J and Hooijschuur,  Peter and Epping,  J\"{o}rn P and de Goede,  Michiel and Kassenberg,  Ben and Venderbosch,  Pim and Toebes,  Chris and van den Vlekkert,  Hans and Pinkse,  Pepijn W H and Renema,  Jelmer J},
  year = {2021},
  pages = {035002}
}

@article{wang09,
  title = {Operator fidelity susceptibility: An indicator of quantum criticality},
  author = {Wang, Xiaoguang and Sun, Zhe and Wang, Z. D.},
  journal = {Phys. Rev. A},
  volume = {79},
  issue = {1},
  pages = {012105},
  numpages = {5},
  year = {2009},
  publisher = {American Physical Society},
  doi = {10.1103/PhysRevA.79.012105},
  url = {https://link.aps.org/doi/10.1103/PhysRevA.79.012105}
}

@article{Clements16,
author = {William R. Clements and Peter C. Humphreys and Benjamin J. Metcalf and W. Steven Kolthammer and Ian A. Walmsley},
journal = {Optica},
number = {12},
pages = {1460--1465},
publisher = {Optica Publishing Group},
title = {Optimal design for universal multiport interferometers},
volume = {3},
year = {2016},
url = {https://opg.optica.org/optica/abstract.cfm?URI=optica-3-12-1460},
doi = {10.1364/OPTICA.3.001460},
}

@misc{Ammara2025arx,
  doi = {10.48550/arxiv.2508.13318},
  url = {https://arxiv.org/abs/2508.13318},
  author = {Ammara,  Ammara and Potoček,  Václav and Štefaňák,  Martin and Pepe,  Francesco V.},
  title = {Quantum Walk on a Line with Absorbing Boundaries},
  publisher = {arXiv},
  year = {2025},
  copyright = {Creative Commons Attribution 4.0 International}
}

@article{Nitsche2018sciad,
  title = {Probing measurement-induced effects in quantum walks via recurrence},
  volume = {4},
  ISSN = {2375-2548},
  url = {http://dx.doi.org/10.1126/sciadv.aar6444},
  DOI = {10.1126/sciadv.aar6444},
  number = {6},
  journal = {Science Advances},
  publisher = {American Association for the Advancement of Science (AAAS)},
  author = {Nitsche,  Thomas and Barkhofen,  Sonja and Kruse,  Regina and Sansoni,  Linda and Štefaňák,  Martin and Gábris,  Aurél and Potoček,  Václav and Kiss,  Tamás and Jex,  Igor and Silberhorn,  Christine},
  year = {2018},
}

@ARTICLE{Stefanak2008,
  title     = "Recurrence and p{\'o}lya number of quantum walks",
  author    = "Stefan{\'a}k, M and Jex, I and Kiss, T",
  journal   = "Phys. Rev. Lett.",
  publisher = "American Physical Society (APS)",
  volume    =  100,
  number    =  2,
  pages     = "020501",
  year      =  2008,
  copyright = "http://link.aps.org/licenses/aps-default-license",
  language  = "en"
}

@article{Biggerstaff16,
	author = {Biggerstaff, Devon N. and Heilmann, Ren{\'e} and Zecevik, Aidan A. and Gr{\"a}fe, Markus and Broome, Matthew A. and Fedrizzi, Alessandro and Nolte, Stefan and Szameit, Alexander and White, Andrew G. and Kassal, Ivan},
	journal = {Nature Comm.},
	number = {1},
	pages = {11282},
	title = {Enhancing coherent transport in a photonic network using controllable decoherence},
	volume = {7},
	year = {2016},
    doi={10.1038/ncomms11282}}

@article{mohs08jcp,
    author = {Mohseni, Masoud and Rebentrost, Patrick and Lloyd, Seth and Aspuru-Guzik, Alan},
    title = {Environment-assisted quantum walks in photosynthetic energy transfer},
    journal = {The Journal of Chemical Physics},
    volume = {129},
    number = {17},
    pages = {174106},
    year = {2008},
    month = {11},
    issn = {0021-9606},
    doi = {10.1063/1.3002335},
    url = {https://doi.org/10.1063/1.3002335},
    eprint = {},
}

@article{Kendon2003,
  title = {Decoherence can be useful in quantum walks},
  author = {Kendon, Viv and Tregenna, Ben},
  journal = {Phys. Rev. A},
  volume = {67},
  issue = {4},
  pages = {042315},
  numpages = {6},
  year = {2003},
  publisher = {American Physical Society},
  doi = {10.1103/PhysRevA.67.042315},
  url = {https://link.aps.org/doi/10.1103/PhysRevA.67.042315}
}

@article{KENDON_2007, 
title={Decoherence in quantum walks – a review}, 
volume={17}, 
DOI={10.1017/S0960129507006354}, 
number={6}, 
journal={Mathematical Structures in Computer Science}, 
author={Kendon, Viv}, 
year={2007}, 
pages={1169–1220}
}

@Article{Verstraete2009,
author={Verstraete, Frank
and Wolf, Michael M.
and Ignacio Cirac, J.},
title={Quantum computation and quantum-state engineering driven by dissipation},
journal={Nature Physics},
year={2009},
day={01},
volume={5},
number={9},
pages={633-636},
issn={1745-2481},
doi={10.1038/nphys1342},
url={https://doi.org/10.1038/nphys1342}
}

@Article{Engel2007,
author={Engel, Gregory S.
and Calhoun, Tessa R.
and Read, Elizabeth L.
and Ahn, Tae-Kyu
and Man{\v{c}}al, Tom{\'a}{\v{s}}
and Cheng, Yuan-Chung
and Blankenship, Robert E.
and Fleming, Graham R.},
title={Evidence for wavelike energy transfer through quantum coherence in photosynthetic systems},
journal={Nature},
year={2007},
day={01},
volume={446},
number={7137},
pages={782-786},
issn={1476-4687},
doi={10.1038/nature05678},
url={https://doi.org/10.1038/nature05678}
}

@article{Kadian2021,
title = {Quantum walk and its application domains: A systematic review},
journal = {Computer Science Review},
volume = {41},
pages = {100419},
year = {2021},
issn = {1574-0137},
doi = {https://doi.org/10.1016/j.cosrev.2021.100419},
url = {https://www.sciencedirect.com/science/article/pii/S1574013721000599},
author = {Karuna Kadian and Sunita Garhwal and Ajay Kumar}
}

@article{Bach2004,
title = {One-dimensional quantum walks with absorbing boundaries},
journal = {Journal of Computer and System Sciences},
volume = {69},
number = {4},
pages = {562-592},
year = {2004},
issn = {0022-0000},
doi = {https://doi.org/10.1016/j.jcss.2004.03.005},
url = {https://www.sciencedirect.com/science/article/pii/S0022000004000376},
author = {Eric Bach and Susan Coppersmith and Marcel Paz Goldschen and Robert Joynt and John Watrous}
}

@article{Kulinski2020,
  title = {Conditional probability distributions of finite absorbing quantum walks},
  author = {Kuklinski, Parker},
  journal = {Phys. Rev. A},
  volume = {101},
  issue = {3},
  pages = {032309},
  numpages = {13},
  year = {2020},
  publisher = {American Physical Society},
  doi = {10.1103/PhysRevA.101.032309},
  url = {https://link.aps.org/doi/10.1103/PhysRevA.101.032309}
}

@Article{Wang2016,
author={Wang, Kun
and Wu, Nan
and Kuklinski, Parker
and Xu, Ping
and Hu, Haixing
and Song, Fangmin},
title={Grover walks on a line with absorbing boundaries},
journal={Quantum Information Processing},
year={2016},
day={01},
volume={15},
number={9},
pages={3573-3597},
issn={1573-1332},
doi={10.1007/s11128-016-1353-5},
url={https://doi.org/10.1007/s11128-016-1353-5}
}

@article{Pegoraro2023,
doi = {10.1088/1402-4896/acbcaa},
url = {https://doi.org/10.1088/1402-4896/acbcaa},
year = {2023},
publisher = {IOP Publishing},
volume = {98},
number = {3},
pages = {034005},
author = {Pegoraro, Federico and Held, Philip and Barkhofen, Sonja and Brecht, Benjamin and Silberhorn, Christine},
title = {Dynamic conditioning of two particle discrete-time quantum walks},
journal = {Physica Scripta}
}

@book{Wang2013_book,
  title={Physical implementation of quantum walks},
  author={Wang, Jingbo and Manouchehri, Kia},
  volume={10},
  year={2013},
  publisher={Springer}
}

@Article{Venegas2012,
author={Venegas-Andraca, Salvador El{\'i}as},
title={Quantum walks: a comprehensive review},
journal={Quantum Information Processing},
year={2012},
day={01},
volume={11},
number={5},
pages={1015-1106},
issn={1573-1332},
doi={10.1007/s11128-012-0432-5},
url={https://doi.org/10.1007/s11128-012-0432-5}
}

\end{document}